\documentclass[%
reprint,
superscriptaddress,
bibnotes,
 amsmath,amssymb,
prb,
]{revtex4-1}

\usepackage{color}
\usepackage{soul}
\usepackage{graphicx}
\usepackage{dcolumn}
\usepackage{bm}
\usepackage{hyperref}

\begin{document}

\preprint{APS/123-QED}

\title{Magnetic diffuse scattering in artificial kagome spin ice}

\author{Oles Sendetskyi}
 \affiliation{%
Laboratory for Mesoscopic Systems, Department of Materials, ETH Zurich, 8093 Zurich, Switzerland
}%
\affiliation{%
Laboratory for Micro- and Nanotechnology, Paul Scherrer Institute, 5232 Villigen PSI, Switzerland
}%

\author{Luca Anghinolfi}
\affiliation{%
Laboratory for Mesoscopic Systems, Department of Materials, ETH Zurich, 8093 Zurich, Switzerland
}%
\affiliation{%
Laboratory for Micro- and Nanotechnology, Paul Scherrer Institute, 5232 Villigen PSI, Switzerland
}%
\affiliation{%
Laboratory for Neutron Scattering and Imaging, Paul Scherrer Institute, 5232 Villigen PSI, Switzerland
}%

\author{Valerio Scagnoli}
\affiliation{%
Laboratory for Mesoscopic Systems, Department of Materials, ETH Zurich, 8093 Zurich, Switzerland
}%
\affiliation{%
Laboratory for Micro- and Nanotechnology, Paul Scherrer Institute, 5232 Villigen PSI, Switzerland
}%

\author{Gunnar M$\mathrm{\ddot{o}}$ller}
\email{G.Moller@kent.ac.uk}
\thanks{\\Present address: School of Physical Sciences, University of Kent, Canterbury CT2 7NH, United Kingdom}
\affiliation{%
TCM Group, Cavendish Laboratory, University of Cambridge, Cambridge CB3 0HE, United Kingdom
}%

\author{Na$\mathrm{\ddot{e}}$mi Leo}
\affiliation{%
Laboratory for Mesoscopic Systems, Department of Materials, ETH Zurich, 8093 Zurich, Switzerland
}%
\affiliation{%
Laboratory for Micro- and Nanotechnology, Paul Scherrer Institute, 5232 Villigen PSI, Switzerland
}%

\author{Aurora Alberca}
\affiliation{%
Swiss Light Source, Paul Scherrer Institute, 5232 Villigen PSI, Switzerland
}%

\author{Joachim Kohlbrecher}
\affiliation{%
Laboratory for Neutron Scattering and Imaging, Paul Scherrer Institute, 5232 Villigen PSI, Switzerland
}%

\author{Jan L$\mathrm{\ddot{u}}$ning}
\affiliation{%
Sorbonne Universit\'{e}s, UPMC Univ Paris 06, UMR 7614, LCPMR, 75005 Paris, France
}%
\affiliation{%
CNRS, UMR 7614, LCPMR, 75005 Paris, France
}%

\author{Urs Staub}
\affiliation{%
Swiss Light Source, Paul Scherrer Institute, 5232 Villigen PSI, Switzerland
}%

\author{Laura Jane Heyderman}
\email{laura.heyderman@psi.ch}
\affiliation{%
Laboratory for Mesoscopic Systems, Department of Materials, ETH Zurich, 8093 Zurich, Switzerland
}%
\affiliation{%
Laboratory for Micro- and Nanotechnology, Paul Scherrer Institute, 5232 Villigen PSI, Switzerland
}%

\date{\today}

\begin{abstract}
The study of magnetic correlations in dipolar-coupled nanomagnet systems with synchrotron \mbox{X-ray} scattering provides a means to uncover emergent phenomena and exotic phases, in particular in systems with thermally active magnetic moments. From the diffuse signal of soft \mbox{X-ray} resonant magnetic scattering, we have measured magnetic correlations in a highly dynamic artificial kagome spin ice with sub-70~nm Permalloy nanomagnets. On comparing experimental scattering patterns with Monte Carlo simulations based on a needle-dipole model, we conclude that kagome ice I phase correlations exist in our experimental system even in the presence of moment fluctuations, which is analogous to bulk spin ice and spin liquid behavior. In addition, we describe the emergence of quasi-pinch points in the magnetic diffuse scattering in the kagome ice I phase. These quasi-pinch points bear similarities to the fully developed pinch points with singularities of a magnetic Coulomb phase, and continually evolve into the latter on lowering the temperature. The possibility to measure magnetic diffuse scattering with soft X-rays opens the way to study magnetic correlations in a variety of nanomagnetic systems.
\end{abstract}

\pacs{Valid PACS appear here}
\maketitle


\section{\label{sec:introduction}Introduction}

Artificial kagome spin ice is a well-known geometrically frustrated two-dimensional magnetic system.\cite{heyderman2013,nisoli2013} It consists of elongated ferromagnetic nanomagnets placed on the nodes of a kagome lattice, or equivalently on the bonds of the hexagonal lattice (see Fig.~\hyperref[fig:figure1]{1}), and coupled via their dipolar magnetic fields. Most of the experimental studies on artificial spin ice have been performed on static systems\cite{wang2006, tanaka2006, ke2008} or systems with slow magnetization dynamics\cite{farhan2013} using imaging techniques such as magnetic force microscopy (MFM) and X-ray photoemission electron microscopy (PEEM). In recent years, however, thermally active artificial spin systems have gained a considerable interest\cite{kapaklis2012} and provided a successful route to reach the low-energy magnetic states of artificial spin ice.\cite{farhan2013, farhanprl2013, anghinolfi2015,kapaklis2014,morgan2011} Additionally, analysis of the thermal behavior of nanomagnetic systems is particularly important for understanding the limitations of future spintronic devices.\cite{chumak2015} However, as the dynamics of artificial spin systems gets faster, observations using traditional microscopy techniques become limited by their temporal resolution (approximately 1~s for PEEM), so that the magnetic correlations of the systems with rapidly fluctuating moments cannot be probed. Therefore, complementary techniques are needed to study magnetic correlations in systems with faster fluctuation timescales in order to gain information about the magnetic phases over a broad temperature range. This is particularly of interest for thermally-induced magnetization dynamics,\cite{farhan2013,farhan2014} order-disorder transitions\cite{moller2009,chern2011} and spin-wave excitations in magnonic crystals.\cite{krawczyk2014,chumak2015}

Here we show that soft X-ray resonant magnetic scattering (SXRMS) is a highly sensitive momentum-resolved technique for studying magnetic correlations in mesoscopic systems with fast magnetization dynamics. Field-driven studies of athermal systems using this technique have already been reported, with measurements of the X-ray magnetic circular dichroism (XMCD) signals at the Bragg peaks.\cite{morgan2012,perron2013,arnalds2012}  Such measurements can directly give information about the ordering in the sample or the net magnetic moment. However, systems without long-range order, but rather short-range correlations, are characterized by \textit{diffuse} scattering with a relatively weak XMCD signal.

In the present work, we focus our attention on a highly dynamic regime of thermally-active artificial kagome spin ice. The magnetic correlations in this regime have not yet been explored due to difficulties in capturing the weak magnetic diffuse signal, which we overcome by masking out the Bragg peaks. Comparing experimental scattering patterns with the patterns calculated from Monte Carlo simulations, we observe the emergence of quasi-pinch points in the kagome ice I phase of artificial kagome spin ice, and explain their relation to the pinch-point singularities in both in-field\cite{tabata2006, fennell2007} and zero-field spin ice pyrochlores.\cite{fennell2009} While we currently cannot access the kagome ice II phase experimentally, we develop a theoretical understanding of how genuine pinch points in the kagome ice II phase emerge smoothly from precursors or quasi-pinch points in the structure factor of the kagome ice I phase by virtue of breaking the $\mathbb{Z}_2$ symmetry associated with magnetic charge order.\cite{moller2009, brooksbartlett2014, chern2011} These sharp pinch points in the magnetic structure factor\cite{brooksbartlett2014} are characteristic features of the Magnetic Coulomb phase.

Our experimental data resolving X-ray magnetic diffuse scattering in highly dynamic artificial kagome spin ice confirms the existence of kagome ice I phase correlations and thus provides convincing evidence for the picture of a classical spin-liquid that thermally samples the large manifold of near degenerate spin-ice states.

\section{\label{sec:methods}Methods}

\subsection{\label{sec:samples}Sample preparation}

\begin{figure}[b]
\includegraphics[width=70mm]{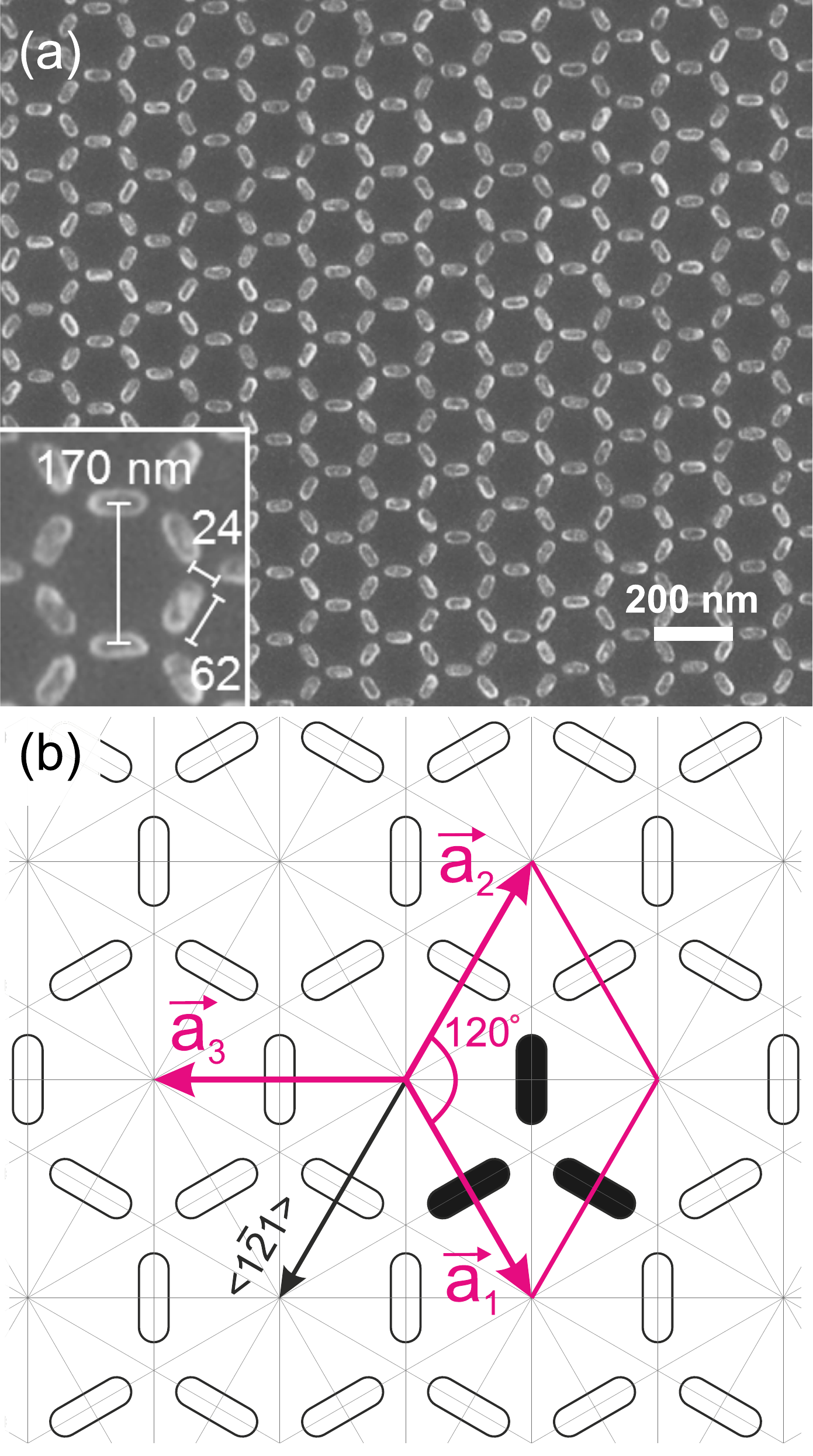}
\label{fig:figure1}
\caption{(a) Scanning electron microscopy image of our artificial kagome spin ice sample with sub-70~nm nanomagnets. A magnified image of a single kagome ring is given in the inset with the lattice and nanomagnet dimensions in nm. (b) Schematic illustration of the nanomagnets on the kagome lattice with the structural unit cell marked in pink, which belongs to $p6mm$ plane group. $\bf{a_1}$, $\bf{a_2}$ and $\bf{a_3}$ are the structural lattice unit vectors. Thin gray lines represent mirror planes and serve as a guide for the eye.  The experimental scattering plane is parallel to the crystallographic direction $\textless$1\={2}1$\textgreater$ indicated by the black arrow. There are three nanomagnets per unit cell, indicated by the three nanomagnets shaded in black.}
\end{figure}

Our 2$\times$2~mm$^2$ arrays of nanomagnets, that are superparamagnetic at room temperature, were produced using electron beam lithography. The nanomagnets have an elongated shape with a length of 62~nm and a width of 24~nm, and are placed on the nodes of a two-dimensional kagome lattice with lattice vectors of length 170~nm (see inset of Fig.~\hyperref[fig:figure1]{1(a)}). A 70~nm-thick polymethylmethacrylate (PMMA) layer was spin-coated on a Si~(100) substrate. The patterns were exposed in the resist using a Vistec EBPG electron beam writer operated at 100~keV accelerating voltage. After development, a  5~nm-thick Permalloy (Ni$_{80}$Fe$_{20}$) film was deposited by thermal evaporation and capped with about 3~nm of Al to prevent oxidation. The remaining resist with unwanted metallic material was removed in acetone by ultrasound-assisted lift-off.

\subsection{\label{sec:simulations}Monte Carlo simulations}

To model the behavior of artificial kagome spin ice, the Monte-Carlo method was used. We have generated magnetic moment configurations for a kagome lattice with Ising degrees of freedom by thermal Monte-Carlo sampling of an array of 24$\times$24 unit cells using the implementation of M$\mathrm{\ddot{o}}$ller and Moessner.\cite{moller2006, moller2009} Each nanomagnet is modeled as a needle dipole with uniform magnetic moment density $\vec{\mu}/l$ along the length $l$ of the nanomagnet,\cite{moller2006} and we use the value of $l/a=0.6$ corresponding to our sample geometry throughout this paper, where $a$ is the distance between the vertices of the kagome lattice or the bond length of the parent hexagonal lattice. The interaction between such needle dipoles is equal to the magnetostatic energy of pairs of magnetic charges $\pm{q}$ at the tips of each dipole, and the value of the Ising spin defines which end hosts the positive charge. This gives rise to an effective description of magnetic configurations in the system in terms of the total magnetic charge $Q$ at a vertex, i.e. the sum of charges at the three dipole tips closest to it (a detailed picture of the energetics in the system can be obtained in a multipole expansion of the available charge configurations).\cite{moller2009} Using this representation, the paramagnetic phase is fully disordered allowing all possible random configurations with $Q=\pm{q}$ or $Q=\pm3q$ at each vertex. In the kagome ice I phase, the ice rule (``two in, one out'' or vice versa at each vertex) enforces $Q=\pm{q}$, with an exponentially suppressed population of $Q=\pm3q$ excitations upon lowering the temperature. The kagome ice II phase is fully ordered in terms of charges but still carries a macroscopic degeneracy in terms of spin configurations, with a finite entropy per magnetic moment.\cite{moller2009} Finally, the (six-fold degenerate) ground state configuration forms a crystal of `loops' of spins with long range order (LRO) in both charge and moment ordering. For more details about these phases see Ref.~\onlinecite{moller2009}. The sampled configurations of the magnetic moments from the kagome ice I and paramagnetic phases are then used to numerically simulate the SXRMS patterns that are compared with the experimental results.

\subsection{\label{sec:softXray}Soft X-ray resonant magnetic diffuse scattering}

\begin{figure}
\includegraphics[width=86mm]{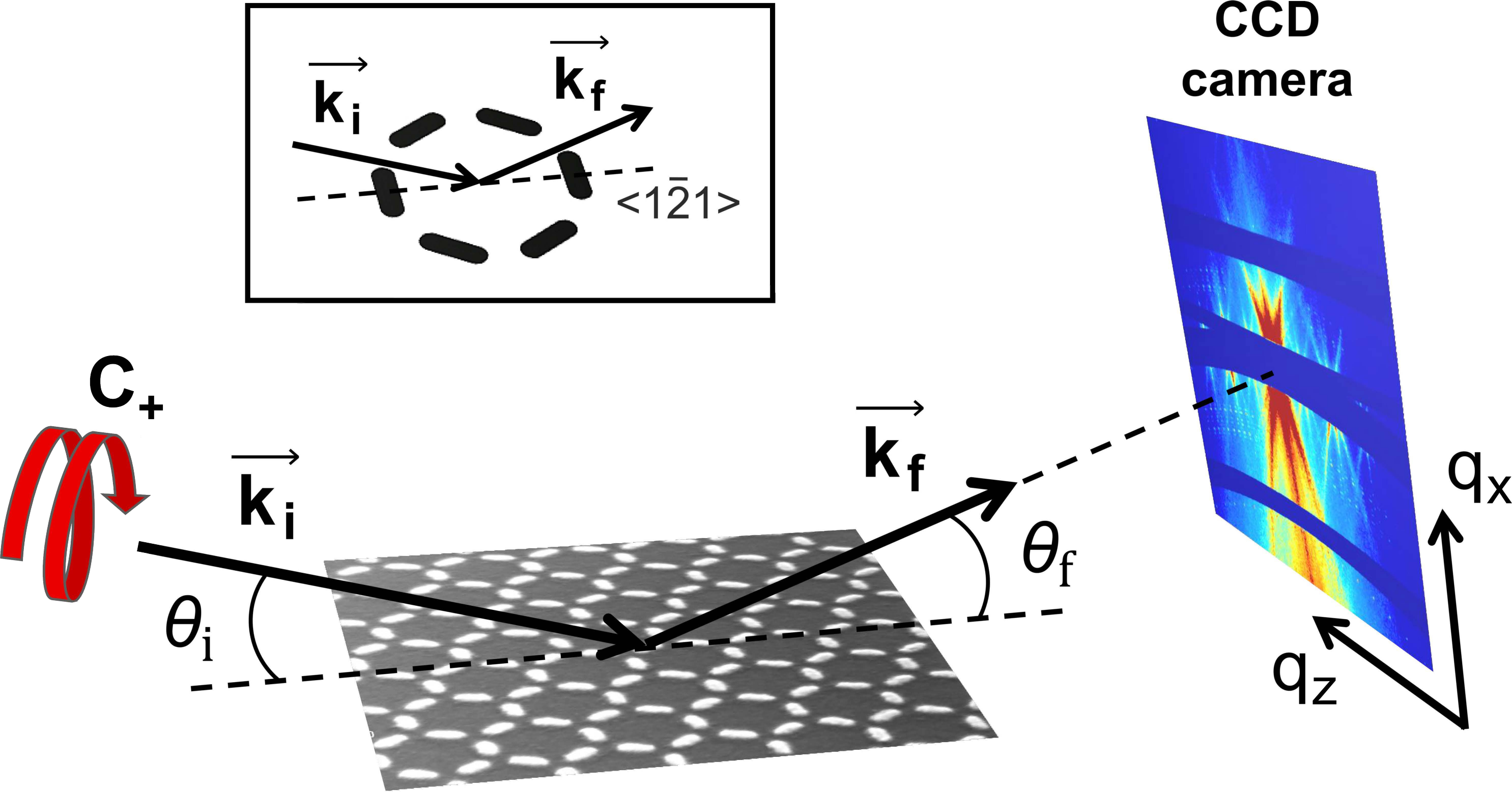} 
       \label{fig:figure2}
       \caption{Experimental geometry for X-ray resonant magnetic scattering with circularly polarized $\mathrm{C_+}$ synchrotron X-rays. $\bf{k_i}$, $\mathrm{\theta_i}$ are the incident X-ray wavevector and angle, and $\bf{k_f}$, $\mathrm{\theta_f}$ are the final wavevector and angle. $\mathrm{q_z}$ and $\mathrm{q_x}$ are the momentum transfers along the z and x directions, respectively. The orientation of the nanomagnets relative to the scattering plane is shown schematically in the inset.}
      \end{figure}

SXRMS experiments were performed using the RESOXS chamber\cite{staub2008} at the SIM beamline\cite{flechsig2010} of the Swiss Light Source, Paul Scherrer  Institut. Experimental  scattering patterns were measured at an X-ray energy of 708~eV corresponding to the Fe L$_3$ edge and at 690~eV (below the edge) to verify that the scattering signal at resonance is of magnetic origin. The off-specular reflection geometry (Fig.~\hyperref[fig:figure2]{2}) ensures sensitivity to the in-plane magnetic moments of the sample. The angle of incidence $\theta_i=8^{\circ}$ is kept constant throughout the experiment. The diffraction patterns were acquired for four seconds with a Princeton Instrument PME charge-coupled device (CCD) camera with 1340$\times$1300~pixels (20$\times$20~$\mu$m$^2$ pixel size).

Using the CCD camera, an extended fraction of reciprocal space can be recorded simultaneously,\cite{perron2013} which is ideally suited for capturing diffuse scattering.  However, the CCD detectors used for soft X-rays have a lower dynamic range compared to the CMOS hybrid photon counting detectors developed for hard X-rays that are used for detecting structural diffuse scattering.\cite{kraft2009} As a result of the low dynamic range, it is much harder to capture diffuse scattering in the soft X-ray scattering experiments, since the diffuse intensity is usually three orders of magnitude lower than the Bragg peak intensity. In addition, the magnetic scattering contribution is usually about two to three orders of magnitude lower than the charge scattering contribution. This results in five to six orders of magnitude difference in intensity between the magnetic diffuse and the structural Bragg scattering, which is beyond the dynamic range of the standard CCD detectors. To separate the two, we therefore placed a custom-made arc-shaped aluminum mask in front of the CCD detector to block the high intensity Bragg peaks and the specular reflection. Thus features in diffuse magnetic scattering with intensities $\approx10^{-6}-10^{-5}$ of the Bragg peak intensities could be resolved. This simple technique opens the possibility to investigate magnetic diffuse scattering, and the associated magnetic correlations, that might otherwise be overlooked in such experiments.

In order to numerically simulate the two-dimensional SXRMS patterns, we make use of kinematic scattering theory as described in Refs.~\onlinecite{blume1985} and \onlinecite{hannon1989}. For more details about the numerical implementation of this theory see Ref.~\onlinecite{van2008} and, in particular, our previous work Ref.~\onlinecite{perron2013}. To directly compare the simulated pattern with the experiment, only the magnetic contribution to the scattering signal was calculated. We assume that the nanomagnets are homogeneous Permalloy particles supporting a single magnetic domain, so we obtain three distinct form factors, one for magnets on each of the three sublattices in the unit cell of the kagome lattice. Improving on the method in Ref. \onlinecite{perron2013}, in this work we also include the scattering geometry and the reflectivity of the sample, which is important to reproduce the shape of the scattering pattern obtained on the 2D detector and the intensity decay in the $q_x$ direction. Taking this into account, we consider a plane wave scattered by the sample and collected at each detector pixel with a momentum transfer $\bf{q}=\bf{k_f}-\bf{k_i}$. The simulated scattering pattern is then corrected for the Fresnel reflectivity from a flat interface, approximated as $(\frac{q_c}{2q_x})^4$ at high angles,\cite{als2011} where ${q_x}$ is the wavevector transfer in the scattering plane and ${q_c}$ is the wavevector transfer at the critical angle.

      \begin{figure}
\includegraphics[width=75mm]{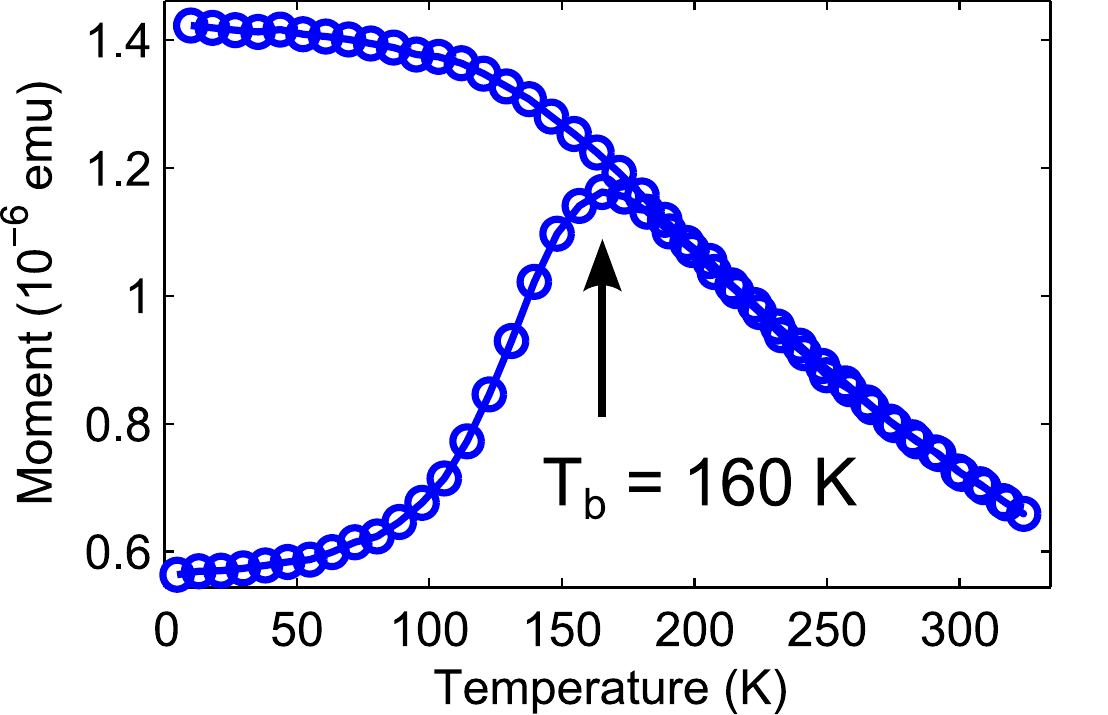} 
       \label{fig:figure3}
       \caption{Zero field-cooled and field-cooled magnetization (ZFC/FC) measurements of our artificial kagome spin ice sample performed with a SQUID magnetometer using an external magnetic field of 10~Oe. The blocking temperature $T_b$~=~160~K is indicated with an arrow. The standard errors are smaller than the data markers.}
      \end{figure}

\section{\label{sec:results}Results and discussion}

\subsection{\label{sec:correlations}Magnetic correlations in highly dynamic artificial kagome spin ice}

In order to investigate the magnetic correlations in the dynamic regime, our samples are designed to achieve an appropriate balance between the magnetic anisotropy energy, the energy associated with the dipolar interaction between the nanomagnets and the thermal energy of the system. The interplay between these energies results in a blocking temperature $T_b$. Tuning the lateral size and thickness of the nanomagnets, as well as the distance between them, we move the blocking temperature to sufficiently low values to ensure fast fluctuations of the magnetic moments, while at the same time keeping the interaction energy between the nanomagnets high enough to preserve collective magnetic behavior. The blocking temperature $T_b$ of the nanomagnets was estimated from zero field-cooled and field-cooled (ZFC/FC) magnetization measurements performed with a Superconducting Quantum Interface Device (SQUID) magnetometer, as shown in Fig.~\hyperref[fig:figure3]{3}. A characteristic peak is found at the blocking temperature $T_b$ that represents a crossover from static to dynamic behavior and is associated with an average energy barrier of the magnetic switching process.\cite{kapaklis2012} For our samples the blocking temperature is around 160~K.

\begin{figure*}[t!]
\includegraphics[width=1\textwidth]{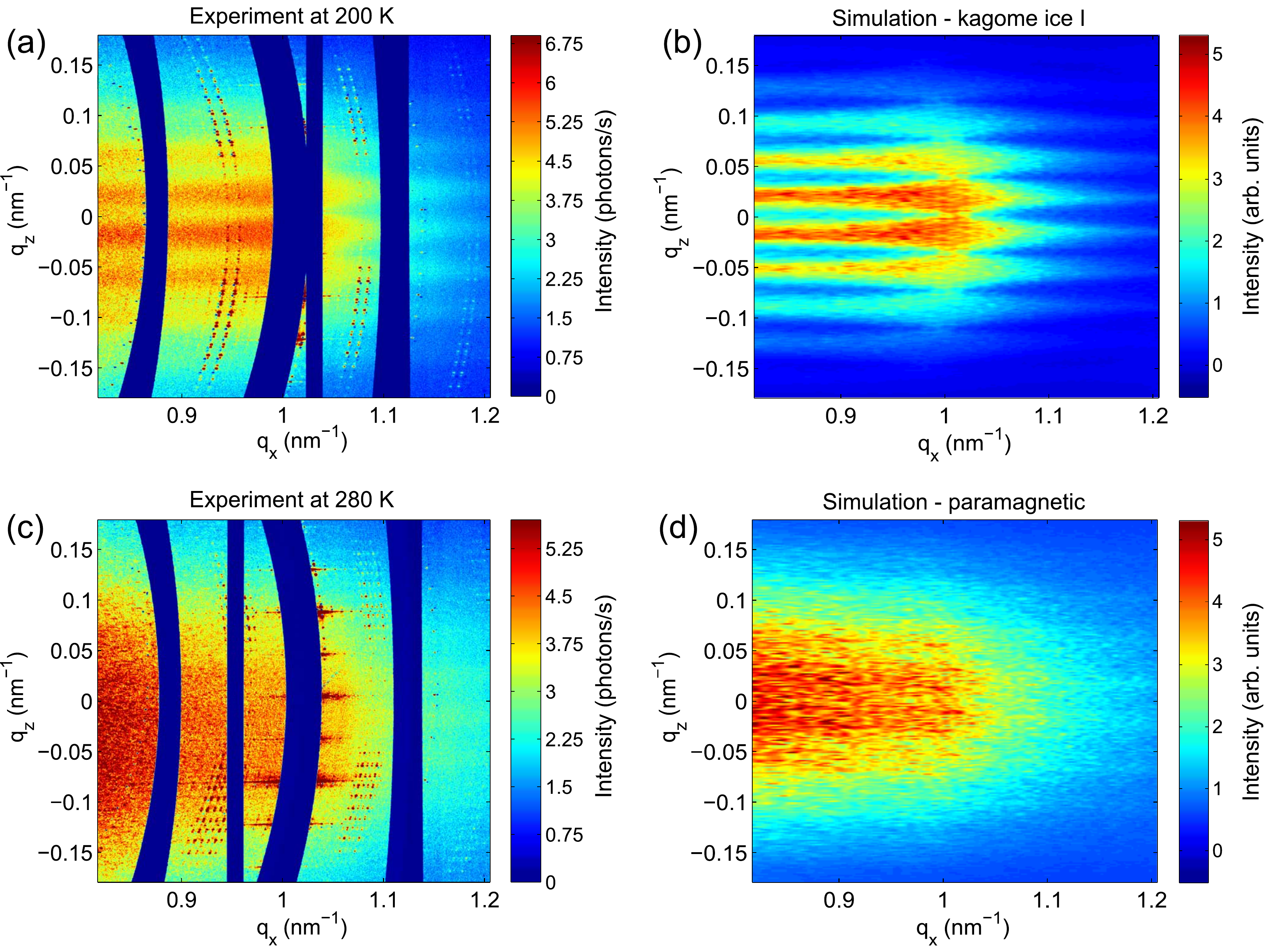} 
       \label{fig:figure4}
       \caption{Comparison between experimental and calculated magnetic diffuse scattering patterns in reflection geometry. (a)~Magnetic scattering pattern measured at 200 K, obtained from the difference between the on-resonance and off-resonance patterns at 708~eV and 690~eV, respectively. (b) Numerical calculation of the resonant magnetic scattering from Monte Carlo configurations in the kagome ice I phase. (c)~Magnetic scattering pattern measured at 280 K, obtained from the difference between the on-resonance and off-resonance patterns at 708~eV and 690~eV, respectively. (d) Numerical calculation of the resonant magnetic scattering from Monte Carlo configurations in the paramagnetic phase. Only the magnetic scattering contribution was simulated. The scattering plane was oriented along the $\textless$1\={2}1$\textgreater$ crystallographic direction (see Figs.~\hyperref[fig:figure1]{1(b)}~and~\hyperref[fig:figure2]{2}). Each experimental pattern is an average over eleven measurements. The dark blue broad stripes on the experimental patterns are the shadows of the aluminum mask, which covers the Bragg peaks and specular reflection. The arcs of closely packed reflections in the experimental patterns arise due to the contamination from higher-order harmonics of the undulator.}
\end{figure*}

In order to probe magnetic correlations in our highly dynamic artificial kagome spin ice, diffuse scattering patterns were measured at several temperatures above the blocking point. Eleven scattering patterns were taken at each temperature to check that the scattering pattern did not change in time and therefore confirm that the sample is thermally equilibrated. To single out the magnetic scattering part (Figs.~\hyperref[fig:figure4]{4(a)}~and~\hyperref[fig:figure4]{4(c)}), we took the difference between on-resonance and off-resonance patterns at X-ray energies of 708 eV and 690 eV, respectively. The off-resonance patterns feature no magnetic diffuse scattering and provide a background intensity. It should be noted here that the arcs of small peaks in the experimental patterns in Figs.~\hyperref[fig:figure4]{4(a)}~and~\hyperref[fig:figure4]{4(c)} arise due to contamination from higher-order harmonics of the undulator. The dark blue broad stripes are the shadows of the aluminum mask, which is used to cover the structural Bragg peaks. The specular reflection is covered by the mask and is estimated to be at $q_x\approx1.03~nm^{-1}$ and $q_z=0~nm^{-1}$. 

At the lower temperature of 200~K, the scattering has bands of intensity oriented along $q_x$, indicating the presence of magnetic correlations (Fig.~\hyperref[fig:figure4]{4(a)}). The stripes become broader near $q_x\approx1.03~nm^{-1}$, forming elongated diamond-shaped features that appear to touch at $q_z=\pm0.037~nm^{-1}$ (the touching point is obscured by the mask). As we will argue below, this feature is related to the formation of ``quasi-pinch points" in the structure factor. Since the temperature of the system is above the blocking temperature $T_b$~=~160~K, the system is dynamic and the magnetic correlations arise from the ice-rule constraints. This constrained collective motion is characteristic of classical spin ices and spin liquids.\cite{bal2010} At the higher temperature of 280~K, the scattering signal becomes uniform indicating that the thermal energy at this temperature is sufficient to break the ice-rule short range correlations of artificial kagome spin ice and it is in a disordered state  (Fig.~\hyperref[fig:figure4]{4(c)}). It should be noted that the intensities of the experimental patterns are slightly asymmetric with higher intensity at negative $q_z$. This is due to a small misalignment of the sample from the $\textless$1\={2}1$\textgreater$ crystallographic direction.

To understand the observed diffuse magnetic scattering patterns, we match our experimental data to predictions based on Monte Carlo simulations. Given a set of spin configurations from a Monte-Carlo run, scattering patterns were obtained using kinematic scattering theory,\cite{blume1985,hannon1989,van2008} as described in the Methods section \ref{sec:methods}. The calculated scattering patterns corresponding to experimental results are shown in Figs.~\hyperref[fig:figure4]{4(b)}~and~\hyperref[fig:figure4]{4(d)}. Experimental data at 200~K visually matches the calculated scattering patterns in the kagome ice I phase. Hence, the scattering pattern reflects ice-like correlations resulting from the ice rule ``two moments in -- one moment out'' or vice versa and vertex charges of $Q =\pm{q}$ are energetically enforced. The experimental data taken at 280~K matches the  scattering patterns calculated from paramagnetic phase configurations. At this temperature, the simulations indicate that the ice rule is frequently violated and vertices with a total charge of $Q =\pm3q$ are common. At both temperatures, we have used an overall prefactor to scale the intensity of numerically simulated patterns from Monte Carlo configurations to match the experimental data. Also, an overall broadening of the experimental data compared with the numerically simulated patterns can be attributed to the limited resolution of the instrument and small variations in the shape of the nanomagnets which produce an additional structural and magnetic diffuse background.

Given the characteristic differences in the scattering within these high- and low-temperature regimes, we can conclude that, from the magnetic diffuse signal obtained with soft X-ray resonant magnetic scattering, we can distinguish magnetic correlations associated with different phases in an artificial spin system.

\subsection{\label{sec:pinch points}Pinch points in the magnetic scattering and the structure factor}

Bulk 3D spin ice realizes a cooperative paramagnet where spins fluctuate in a correlated manner between low energy states.\cite{villain1979} Dipolar correlations are present despite the absence of long-range order and are one of the most interesting features of spin ice systems.\cite{isakov2004,moessner2003} In reciprocal space, their characteristic hallmarks are sharp singularities in the spin structure factor known as bow ties or pinch points. They were initially found in ferroelectrics\cite{skalyo1970,youngblood1981} and more recently, their presence was predicted for many magnetic systems,\cite{isakov2004,moessner2003,henley2005} and they have been observed in both in-field\cite{tabata2006, fennell2007} and zero-field spin ice pyrochlores.\cite{fennell2009} Pinch points reflect the presence of a local divergence free condition of atomic spins, when considering the lattice flux where the local ice rule is obeyed. In the pyrochlore spin ice, the ice rule refers to the local energy minimization condition of spins on the corners of a single tetrahedron with two spins pointing towards and two spins pointing away from the center.\cite{anderson1956}

\begin{figure*}
\includegraphics[width=1\textwidth]{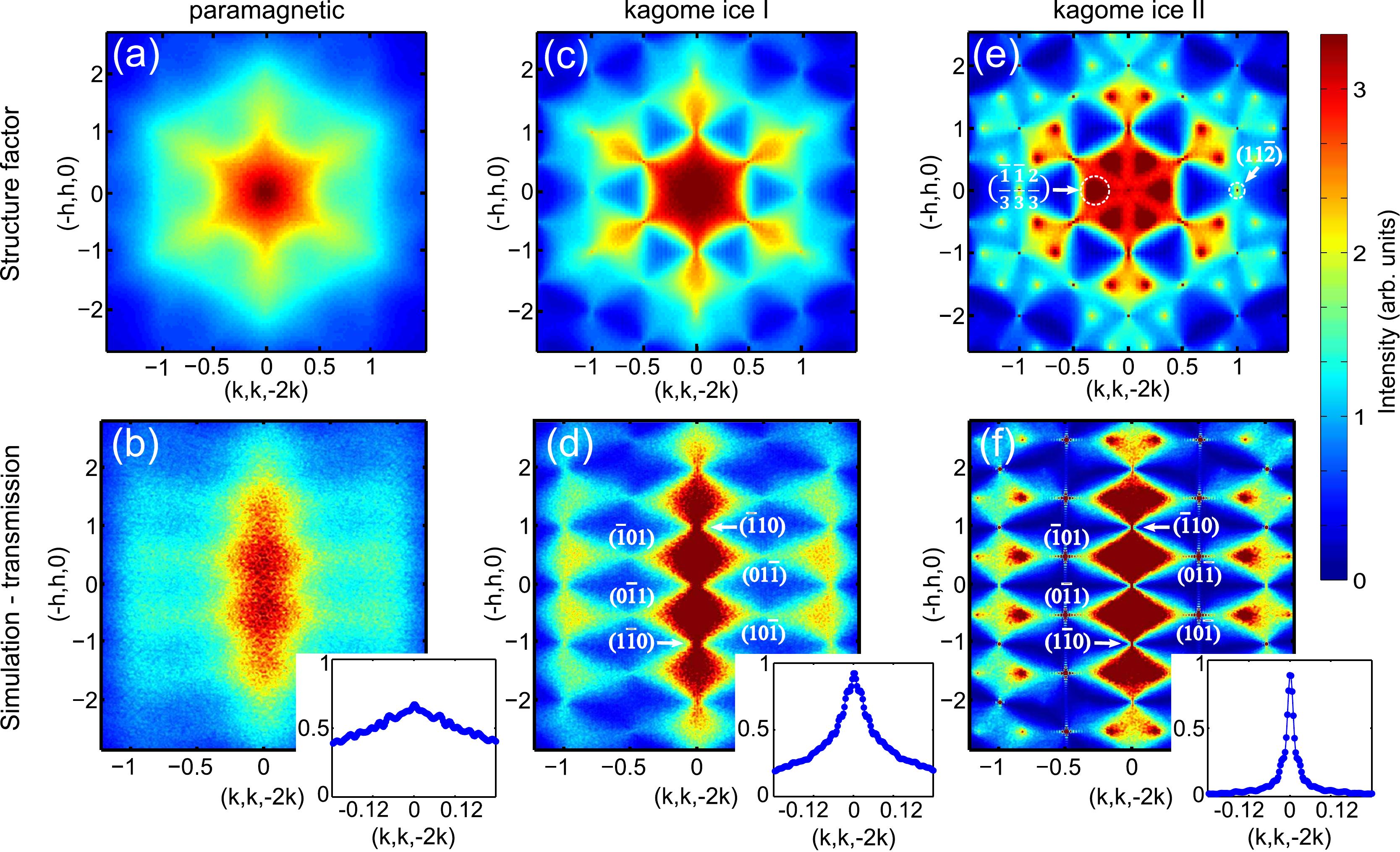} 
       \label{fig:figure5}
       \caption{Numerical calculation of the structure factor (top row) and resonant magnetic scattering in transmission geometry (bottom row). Only the magnetic contribution is shown, as predicted from Monte Carlo simulations. Data for the paramagnetic (a,~b) and kagome ice I phase (c,~d) are based on the same set of Monte Carlo configurations that we used for comparison with the experimental results shown in Fig.~\hyperref[fig:figure4]{4}. For the kagome ice II phase (e,~f) the representative configurations were also taken from our Monte Carlo simulations. Note that magnetic pinch points appear at the same positions as the structural peaks (not shown), e.g. at~(01$\mathrm{\bar{1}}$), (10$\mathrm{\bar{1}}$), (1$\mathrm{\bar{1}}$0), (0$\mathrm{\bar{1}}$1), ($\mathrm{\bar{1}}$01) and ($\mathrm{\bar{1}}$10). Insets in the lower row are the intensity profiles of the pinch point at the ($\mathrm{\bar{1}}$10) position.}
      \end{figure*}

In order to determine whether pinch points may in principle be observed for artificial kagome spin ice, we have calculated the spin structure factor. This is the most familiar quantity and is independent of the particular experimental setup in a neutron or X-ray scattering experiment. As an experimentally accessible signature, we have also calculated the magnetic diffuse scattering in a transmission geometry from our Monte Carlo simulations. Results for these are shown in Fig.~\hyperref[fig:figure5]{5}. For the patterns in the transmission geometry, that we use also to assess the width of the pinch points, we assume an incident angle of 30$^\circ$ and calculate the forward scattering of X-rays. Only the magnetic scattering contribution was calculated. In Figs.~\hyperref[fig:figure5]{5(a)}--\hyperref[fig:figure5]{5(d)}, we show data for the paramagnetic and kagome ice I phases using the same set of Monte Carlo configurations that we used for comparison with the experimental results shown in Fig.~\hyperref[fig:figure4]{4}. In addition, we also show simulation data for the kagome ice II phase in Figs.~\hyperref[fig:figure5]{5(e)}~and~\hyperref[fig:figure5]{5(f)}. In this phase an additional magnetic Bragg contribution appears at the pinch point positions of (01$\mathrm{\bar{1}}$), (10$\mathrm{\bar{1}}$), (0$\mathrm{\bar{1}}$1) and ($\mathrm{\bar{1}}$01)  Fig.~\hyperref[fig:figure5]{5(e)}. However it is absent at the (1$\mathrm{\bar{1}}$0) and ($\mathrm{\bar{1}}$10) positions, which makes the latter ideal for the measurements of the pinch point intensity profiles, see insets in Figs.~\hyperref[fig:figure5]{5(b)},~\hyperref[fig:figure5]{5(d)}~and~\hyperref[fig:figure5]{5(f)}. We also note that the kagome ice II phase can be distinguished from the kagome ice I phase by its partial magnetic order, as also shown in Ref. \onlinecite{brooksbartlett2014}. Evidence of this can be best seen by the magnetic Bragg peaks appearing in kagome ice II phase at, for example, (11$\mathrm{\bar{2}}$) or ($\mathrm{\bar{\frac{1}{3}}}$$\mathrm{\bar{\frac{1}{3}}}$$\frac{2}{3}$) and equivalent positions [see Fig.~\hyperref[fig:figure5]{5(e)}]. It should also be noted that, for the scattering patterns of Fig.~\hyperref[fig:figure5]{5}, we have used the crystallographic conventions for the hexagonal planar group $p6mm$\cite{burzlaff2006, steurer2009} with the unit cell defined in Fig.~\hyperref[fig:figure1]{1(b)}. To describe this planar hexagonal lattice, we employ $hki$ indices, with the $i$ index being redundant. For the native kagome unit cell, the positions of the pinch points are described by $hki$ indices that have $integer$ values. In previous works \cite{fennell2007, fennell2009, brooksbartlett2014} that considered the three-dimensional pyrochlore unit cell, the pinch point positions are described by $fractional$ $hkl$ indices. 

To show the evolution of the correlations across the different phases, let us first consider the structure factor. Upon lowering the temperature from the paramagnetic phase, where the scattering signal is uniform  Fig.~\hyperref[fig:figure5]{5(a)}, characteristic regions of high- and low-intensity scattering start to emerge and form quasi-pinch points in the kagome ice I phase (Fig.~\hyperref[fig:figure5]{5(c)}). These subsequently evolve into sharp, singular features in the kagome ice II phase (Fig.~\hyperref[fig:figure5]{5(e)}). An analogous evolution of the of the pinch points is seen in the transmission geometry calculations [see Figs.~\hyperref[fig:figure5]{5(b)},~\hyperref[fig:figure5]{5(d)},~\hyperref[fig:figure5]{5(f)} and insets]. Finally, in the reflection geometry of our experiments, we now recognize that the previously described touching of elongated diamond-shaped features at regular points in reciprocal space (Figs.~\hyperref[fig:figure4]{4(a)}~and~\hyperref[fig:figure4]{4(b)}), should also evolve into sharper and sharper features that correspond to the appearance of the pinch points in Fig.~\hyperref[fig:figure5]{5}. As already noted, the magnetic pinch points emerge at the positions of the structural Bragg peaks (not shown), some of which we have indicated in Figs.~\hyperref[fig:figure5]{5(d)}~and~\hyperref[fig:figure5]{5(f)}, for example at (01$\mathrm{\bar{1}}$), (10$\mathrm{\bar{1}}$), (1$\mathrm{\bar{1}}$0), (0$\mathrm{\bar{1}}$1), ($\mathrm{\bar{1}}$01) and ($\mathrm{\bar{1}}$10) at the edges of the first Brillouin zone.

We now explain these observations in more detail. The emergence of pinch points requires the presence of a local conservation law for lattice fluxes,\cite{youngblood1981, huse2003, isakov2004, moessner2003, henley2010} which is one of the defining features of a Coulomb phase. In particular, the observation of pinch points in neutron scattering patterns has provided a direct experimental evidence for a field-induced and zero-field Coulomb phase in pyrochlore magnets. \cite{tabata2006,fennell2007,fennell2009,isakov2004,moessner2003} However, the trivalent vertices of the planar kagome lattice cannot inherently realize conservation of flux, and only allow configurations with a local lattice divergence of absolute value $\geq 1$. Nevertheless, the charge-ordered kagome ice II phase can be shown to be a Coulomb phase, due to the fact that it spontaneously breaks the $\mathbb{Z}_2$ sublattice symmetry due to the formation of magnetic charge order.\cite{moller2009, chern2011} In this symmetry broken phase, the allowed spin configurations are identical to the constraints found on the kagome planes of the three-dimensional pyrochlore spin ice with a magnetic field along the [111] direction that was experimentally investigated in Refs.~\onlinecite{fennell2007,tabata2006}. In this 3D situation, the sublattice symmetry of the (buckled) kagome planes of the pyrochlore lattice is inherently broken, as one sublattice features out-of-plane spins pointing along [111] direction, while the out-of-plane spins on the other sublattice point towards the negative [111] direction. In the presence of a field along the [111] axis, the flux along that direction is fixed and vertices in the kagome planes are forced to have alternating minority spins (pointing towards or away from the center of a tetrahedron) \cite{udagawa2002}, which amounts to an explicit breaking of the sublattice symmetry. The local flux conservation is fulfilled for the low-energy configurations in this three-dimensional setting and, by extension, the same is true for the kagome ice II phase. However, this 3D picture requires the global sublattice symmetry to be explicitly broken, and it would be desirable to develop a more local picture that can accommodate the spontaneous symmetry breaking of the ice I to ice II transition. Indeed, such a viewpoint can be developed on the basis of a dimer mapping.\cite{huse2003,misguich2002} 

Let us consider the mapping from oriented hard-core dimers to a divergence-free configuration of fluxes on the links of the lattice by Huse et al.\cite{huse2003}. Indeed, due to the $\mathbb{Z}_2$ sublattice symmetry breaking of magnetic charge order, the kagome ice II phase allows a unique mapping of spin configurations to oriented hard-core dimers on the kagome lattice:\cite{moller2009, chern2011,udagawa2002} At each vertex, there is precisely one minority charge, which connects to the minority charge of one adjacent vertex, and uniquely identifies the corresponding link as carrying a dimer oriented from the negative minority charge to the positive one, while the other two links are empty.\cite{moller2009, chern2011} Following Huse et al.,\cite{huse2003} we can then define the lattice flux carried by a link in terms of the occupation number $n_i(\mathbf{r})$ of dimers on the link $\mathbf{r}\to\mathbf{r}+\mathbf{e}_i$ as

\begin{equation}
\label{eq:LatticeFlux}
\mathbf{B}_i(\mathbf{r})= Q(\mathbf{r})\left(n_i(\mathbf{r})-\frac{1}{3}\right),
\end{equation}

using the fact that the total monopole charge $Q(\mathbf{r})=\pm1$ has long-range order in the kagome ice II phase and thus sets the orientation of the fluxes for the respective sublattices. Note that this lattice flux has twice the magnitude on the links carrying dimers compared with the flux on the empty ones. Therefore, one can picture a pair of fluxes entering one vertex, being carried to an adjacent vertex along the dimer, and re-emerging as two separate fluxes from there. Clearly, such ``flux dimers'' satisfy flux conservation. Thus, as in the case of the square and cubic lattices of Huse et al.,\cite{huse2003} the lattice divergence vanishes, and the field can be cast into the habitual language of a lattice gauge theory and forms a Coulomb phase. As the kagome ice II phase is a genuine Coulomb phase with a local conservation law, its structure factor is predicted to feature sharp singularities at the pinch points.\cite{brooksbartlett2014} This is shown by the pinch point intensity profile in the inset of Fig.~\hyperref[fig:figure5]{5(f)}.

To proceed with our discussion, we note that spin configurations can be uniquely mapped to a dimer covering only in the monopole charge ordered kagome ice II phase. By contrast, the kagome ice I phase has no long range charge order and thus must violate the divergence-free condition. Consequently, it does not allow for uniquely defined flux dimers as in Eq.~\hyperref[eq:LatticeFlux]{(1)}.\footnote{More general flux dimers could be defined, such that the dimer carries an effective flux of zero, and the two adjacent vertices have one spin in- and out- each. However, one cannot assign such dimers in a unique fashion in either of the ice I or ice II phases.} Nevertheless, there will be charge ordered clusters in the kagome ice I phase so that, while a global dimer mapping does not exist, it is still possible to locally identify dimers within each charge-ordered cluster. As the structure factor is expressed in reciprocal space, and can be thought of as a spatial average over the system, the overall structure encoded in the local conservation law of the low-temperature ordered phase is therefore also observed in the kagome ice I phase. Additionally, the larger number of vertices that are not covered by flux dimers gives rise to the broadening of the pinch-points.

Indeed, a broadening of the pinch points can be seen in our calculations of the structure factor and of the resonant magnetic scattering in a transmission geometry for the kagome ice I phase (Fig.~\hyperref[fig:figure5]{5}). Comparing the calculations from the different phases in Fig.~\hyperref[fig:figure5]{5}, we note that the quasi-pinch-points in the scattering profiles of the kagome ice I phase are very similar to the fully developed pinch points of kagome ice II phase. However, following the arguments above, no sharp singularity can exist at the center of these quasi-pinch points. Finally in the Paramagnetic phase there is a random distribution of $Q =\pm{q}$ and $Q =\pm3q$ charges. Therefore a smoothly varying diffuse signal is established, as can be seen in Figs.~\hyperref[fig:figure5]{5(a)}~and~\hyperref[fig:figure5]{5(b)}.
 
It should be noted that, in general, the experimental detection of the pinch points in a transmission geometry is extremely challenging: due to the p6mm symmetry of the lattice and the absence of systematic extinctions, structural Bragg peaks are positioned at the same $\bf{q}$ values as the pinch points, thus overlapping with them. Furthermore, a soft X-ray detector with the required dynamic range to capture both strong Bragg scattering and a weaker magnetic diffuse scattering is not commonly available. Despite these difficulties, our results show that it is possible to qualitatively distinguish the different phases of artificial kagome spin ice and capture features that are directly related to pinch points in the reflection geometry employed in our set-up.

Given these insights, we conclude that our experimental data provides the first evidence of the emergence of pinch-point scattering in artificial kagome spin ice. These correlations are highly distinct from those found in the high-temperature paramagnetic phase, and present direct evidence that the ice-rules are obeyed, reflecting the spin dynamics that minimize the dominant nearest neighbor interaction terms of the needle dipole Hamiltonian on the kagome lattice.

\section{\label{sec:conclusions}Conclusions}

We have demonstrated that the measurement of diffuse scattering with resonant synchrotron X-rays provides a highly sensitive method for the investigation of short-range correlations in nanomagnet systems. In particular, we have shown that the phases of artificial kagome spin ice can be distinguished and comparison with Monte Carlo simulations confirms the realization of kagome ice I phase magnetic correlations at high moment fluctuation rates. This combination of a highly dynamic system with strong short-range correlations is characteristic of a classical spin liquid (or a cooperative paramagnet). Although the kagome ice I phase is not a genuine Coulomb phase, since the divergence-free condition does not extend throughout the lattice, the features of quasi-pinch points already bear similarities to the fully-developed pinch points with sharp singularities associated with the kagome ice II Coulomb phase. We have used a mapping to oriented dimer coverings to argue that such quasi-pinch points should indeed be expected above the phase transition to the charge ordered kagome ice II phase. 

We conclude that, like in the bulk pyrochlore spin ices, diffuse scattering gives a unique signature of ice-like behavior in artificial spin ices. This technique also promises to cast light onto a number of other problems in nanoscale magnetic systems, for example to understand the ordering processes during magnetic self-assembly of nanoparticles\cite{carbone2011} or the formation of skyrmion lattices.\cite{muhlbauer2009} This highly sensitive method will therefore be an important tool for the discovery of novel physics in two-dimensional systems beyond artificial spin ice.

\begin{acknowledgments}
We are grateful to V.~Guzenko, A.~Weber, T.~Neiger, M.~Horisberger and E.~Deckardt for their support in the sample preparation, N.~S.~Bingham for his help with experiments, P.~Schifferle for technical support at the X11MA beamline, and T.~Fennell and R.~Moessner for helpful discussions. Part of this work was carried out at the X11MA beamline of the Swiss Light Source, Paul Scherrer Institute, Villigen, Switzerland. The first soft X-ray magnetic diffuse scattering patterns were taken at the SEXTANTS beamline of Synchrotron SOLEIL, Saclay, France under the proposal~20120875. The research leading to these results has received funding from the Swiss National Science Foundation, the European Community's Seventh Framework Programme (FP7/2007-2013) -- Grant No.~290605 (COFUND PSI-FELLOW) and the Royal Society -- Grant No.~UF120157.
\end{acknowledgments}



\begin{thebibliography}{48}%
\makeatletter
\providecommand \@ifxundefined [1]{%
 \@ifx{#1\undefined}
}%
\providecommand \@ifnum [1]{%
 \ifnum #1\expandafter \@firstoftwo
 \else \expandafter \@secondoftwo
 \fi
}%
\providecommand \@ifx [1]{%
 \ifx #1\expandafter \@firstoftwo
 \else \expandafter \@secondoftwo
 \fi
}%
\providecommand \natexlab [1]{#1}%
\providecommand \enquote  [1]{``#1''}%
\providecommand \bibnamefont  [1]{#1}%
\providecommand \bibfnamefont [1]{#1}%
\providecommand \citenamefont [1]{#1}%
\providecommand \href@noop [0]{\@secondoftwo}%
\providecommand \href [0]{\begingroup \@sanitize@url \@href}%
\providecommand \@href[1]{\@@startlink{#1}\@@href}%
\providecommand \@@href[1]{\endgroup#1\@@endlink}%
\providecommand \@sanitize@url [0]{\catcode `\\12\catcode `\$12\catcode
  `\&12\catcode `\#12\catcode `\^12\catcode `\_12\catcode `\%12\relax}%
\providecommand \@@startlink[1]{}%
\providecommand \@@endlink[0]{}%
\providecommand \url  [0]{\begingroup\@sanitize@url \@url }%
\providecommand \@url [1]{\endgroup\@href {#1}{\urlprefix }}%
\providecommand \urlprefix  [0]{URL }%
\providecommand \Eprint [0]{\href }%
\providecommand \doibase [0]{http://dx.doi.org/}%
\providecommand \selectlanguage [0]{\@gobble}%
\providecommand \bibinfo  [0]{\@secondoftwo}%
\providecommand \bibfield  [0]{\@secondoftwo}%
\providecommand \translation [1]{[#1]}%
\providecommand \BibitemOpen [0]{}%
\providecommand \bibitemStop [0]{}%
\providecommand \bibitemNoStop [0]{.\EOS\space}%
\providecommand \EOS [0]{\spacefactor3000\relax}%
\providecommand \BibitemShut  [1]{\csname bibitem#1\endcsname}%
\let\auto@bib@innerbib\@empty
\bibitem [{\citenamefont {Heyderman}\ and\ \citenamefont
  {Stamps}(2013)}]{heyderman2013}%
  \BibitemOpen
  \bibfield  {author} {\bibinfo {author} {\bibfnamefont {L.~J.}\ \bibnamefont
  {Heyderman}}\ and\ \bibinfo {author} {\bibfnamefont {R.~L.}\ \bibnamefont
  {Stamps}},\ }\href@noop {} {\bibfield  {journal} {\bibinfo  {journal}
  {Journal of Physics: Condensed Matter}\ }\textbf {\bibinfo {volume} {25}},\
  \bibinfo {pages} {363201} (\bibinfo {year} {2013})}\BibitemShut {NoStop}%
\bibitem [{\citenamefont {Nisoli}\ \emph {et~al.}(2013)\citenamefont {Nisoli},
  \citenamefont {Moessner},\ and\ \citenamefont {Schiffer}}]{nisoli2013}%
  \BibitemOpen
  \bibfield  {author} {\bibinfo {author} {\bibfnamefont {C.}~\bibnamefont
  {Nisoli}}, \bibinfo {author} {\bibfnamefont {R.}~\bibnamefont {Moessner}}, \
  and\ \bibinfo {author} {\bibfnamefont {P.}~\bibnamefont {Schiffer}},\
  }\href@noop {} {\bibfield  {journal} {\bibinfo  {journal} {Reviews of Modern
  Physics}\ }\textbf {\bibinfo {volume} {85}},\ \bibinfo {pages} {1473}
  (\bibinfo {year} {2013})}\BibitemShut {NoStop}%
\bibitem [{\citenamefont {Wang}\ \emph {et~al.}(2006)\citenamefont {Wang},
  \citenamefont {Nisoli}, \citenamefont {Freitas}, \citenamefont {Li},
  \citenamefont {McConville}, \citenamefont {Cooley}, \citenamefont {Lund},
  \citenamefont {Samarth}, \citenamefont {Leighton}, \citenamefont {Crespi}
  \emph {et~al.}}]{wang2006}%
  \BibitemOpen
  \bibfield  {author} {\bibinfo {author} {\bibfnamefont {R.~F.}\ \bibnamefont
  {Wang}}, \bibinfo {author} {\bibfnamefont {C.}~\bibnamefont {Nisoli}},
  \bibinfo {author} {\bibfnamefont {R.~S.}\ \bibnamefont {Freitas}}, \bibinfo
  {author} {\bibfnamefont {J.}~\bibnamefont {Li}}, \bibinfo {author}
  {\bibfnamefont {W.}~\bibnamefont {McConville}}, \bibinfo {author}
  {\bibfnamefont {B.~J.}\ \bibnamefont {Cooley}}, \bibinfo {author}
  {\bibfnamefont {M.~S.}\ \bibnamefont {Lund}}, \bibinfo {author}
  {\bibfnamefont {N.}~\bibnamefont {Samarth}}, \bibinfo {author} {\bibfnamefont
  {C.}~\bibnamefont {Leighton}}, \bibinfo {author} {\bibfnamefont {V.~H.}\
  \bibnamefont {Crespi}},  \emph {et~al.},\ }\href@noop {} {\bibfield
  {journal} {\bibinfo  {journal} {Nature}\ }\textbf {\bibinfo {volume} {439}},\
  \bibinfo {pages} {303} (\bibinfo {year} {2006})}\BibitemShut {NoStop}%
\bibitem [{\citenamefont {Tanaka}\ \emph {et~al.}(2006)\citenamefont {Tanaka},
  \citenamefont {Saitoh}, \citenamefont {Miyajima}, \citenamefont {Yamaoka},\
  and\ \citenamefont {Iye}}]{tanaka2006}%
  \BibitemOpen
  \bibfield  {author} {\bibinfo {author} {\bibfnamefont {M.}~\bibnamefont
  {Tanaka}}, \bibinfo {author} {\bibfnamefont {E.}~\bibnamefont {Saitoh}},
  \bibinfo {author} {\bibfnamefont {H.}~\bibnamefont {Miyajima}}, \bibinfo
  {author} {\bibfnamefont {T.}~\bibnamefont {Yamaoka}}, \ and\ \bibinfo
  {author} {\bibfnamefont {Y.}~\bibnamefont {Iye}},\ }\href@noop {} {\bibfield
  {journal} {\bibinfo  {journal} {Physical Review B}\ }\textbf {\bibinfo
  {volume} {73}},\ \bibinfo {pages} {052411} (\bibinfo {year}
  {2006})}\BibitemShut {NoStop}%
\bibitem [{\citenamefont {Ke}\ \emph {et~al.}(2008)\citenamefont {Ke},
  \citenamefont {Li}, \citenamefont {Nisoli}, \citenamefont {Lammert},
  \citenamefont {McConville}, \citenamefont {Wang}, \citenamefont {Crespi},\
  and\ \citenamefont {Schiffer}}]{ke2008}%
  \BibitemOpen
  \bibfield  {author} {\bibinfo {author} {\bibfnamefont {X.}~\bibnamefont
  {Ke}}, \bibinfo {author} {\bibfnamefont {J.}~\bibnamefont {Li}}, \bibinfo
  {author} {\bibfnamefont {C.}~\bibnamefont {Nisoli}}, \bibinfo {author}
  {\bibfnamefont {P.~E.}\ \bibnamefont {Lammert}}, \bibinfo {author}
  {\bibfnamefont {W.}~\bibnamefont {McConville}}, \bibinfo {author}
  {\bibfnamefont {R.~F.}\ \bibnamefont {Wang}}, \bibinfo {author}
  {\bibfnamefont {V.~H.}\ \bibnamefont {Crespi}}, \ and\ \bibinfo {author}
  {\bibfnamefont {P.}~\bibnamefont {Schiffer}},\ }\href@noop {} {\bibfield
  {journal} {\bibinfo  {journal} {Physical Review Letters}\ }\textbf {\bibinfo
  {volume} {101}},\ \bibinfo {pages} {037205} (\bibinfo {year}
  {2008})}\BibitemShut {NoStop}%
\bibitem [{\citenamefont {Farhan}\ \emph
  {et~al.}(2013{\natexlab{a}})\citenamefont {Farhan}, \citenamefont {Derlet},
  \citenamefont {Kleibert}, \citenamefont {Balan}, \citenamefont {Chopdekar},
  \citenamefont {Wyss}, \citenamefont {Anghinolfi}, \citenamefont {Nolting},\
  and\ \citenamefont {Heyderman}}]{farhan2013}%
  \BibitemOpen
  \bibfield  {author} {\bibinfo {author} {\bibfnamefont {A.}~\bibnamefont
  {Farhan}}, \bibinfo {author} {\bibfnamefont {P.~M.}\ \bibnamefont {Derlet}},
  \bibinfo {author} {\bibfnamefont {A.}~\bibnamefont {Kleibert}}, \bibinfo
  {author} {\bibfnamefont {A.}~\bibnamefont {Balan}}, \bibinfo {author}
  {\bibfnamefont {R.~V.}\ \bibnamefont {Chopdekar}}, \bibinfo {author}
  {\bibfnamefont {M.}~\bibnamefont {Wyss}}, \bibinfo {author} {\bibfnamefont
  {L.}~\bibnamefont {Anghinolfi}}, \bibinfo {author} {\bibfnamefont
  {F.}~\bibnamefont {Nolting}}, \ and\ \bibinfo {author} {\bibfnamefont
  {L.~J.}\ \bibnamefont {Heyderman}},\ }\href@noop {} {\bibfield  {journal}
  {\bibinfo  {journal} {Nature Physics}\ }\textbf {\bibinfo {volume} {9}},\
  \bibinfo {pages} {375} (\bibinfo {year} {2013}{\natexlab{a}})}\BibitemShut
  {NoStop}%
\bibitem [{\citenamefont {Kapaklis}\ \emph {et~al.}(2012)\citenamefont
  {Kapaklis}, \citenamefont {Arnalds}, \citenamefont {Harman-Clarke},
  \citenamefont {Papaioannou}, \citenamefont {Karimipour}, \citenamefont
  {Korelis}, \citenamefont {Taroni}, \citenamefont {Holdsworth}, \citenamefont
  {Bramwell},\ and\ \citenamefont {Hj{\"o}rvarsson}}]{kapaklis2012}%
  \BibitemOpen
  \bibfield  {author} {\bibinfo {author} {\bibfnamefont {V.}~\bibnamefont
  {Kapaklis}}, \bibinfo {author} {\bibfnamefont {U.~B.}\ \bibnamefont
  {Arnalds}}, \bibinfo {author} {\bibfnamefont {A.}~\bibnamefont
  {Harman-Clarke}}, \bibinfo {author} {\bibfnamefont {E.~T.}\ \bibnamefont
  {Papaioannou}}, \bibinfo {author} {\bibfnamefont {M.}~\bibnamefont
  {Karimipour}}, \bibinfo {author} {\bibfnamefont {P.}~\bibnamefont {Korelis}},
  \bibinfo {author} {\bibfnamefont {A.}~\bibnamefont {Taroni}}, \bibinfo
  {author} {\bibfnamefont {P.~C.}\ \bibnamefont {Holdsworth}}, \bibinfo
  {author} {\bibfnamefont {S.~T.}\ \bibnamefont {Bramwell}}, \ and\ \bibinfo
  {author} {\bibfnamefont {B.}~\bibnamefont {Hj{\"o}rvarsson}},\ }\href@noop {}
  {\bibfield  {journal} {\bibinfo  {journal} {New Journal of Physics}\ }\textbf
  {\bibinfo {volume} {14}},\ \bibinfo {pages} {035009} (\bibinfo {year}
  {2012})}\BibitemShut {NoStop}%
\bibitem [{\citenamefont {Farhan}\ \emph
  {et~al.}(2013{\natexlab{b}})\citenamefont {Farhan}, \citenamefont {Derlet},
  \citenamefont {Kleibert}, \citenamefont {Balan}, \citenamefont {Chopdekar},
  \citenamefont {Wyss}, \citenamefont {Perron}, \citenamefont {Scholl},
  \citenamefont {Nolting},\ and\ \citenamefont {Heyderman}}]{farhanprl2013}%
  \BibitemOpen
  \bibfield  {author} {\bibinfo {author} {\bibfnamefont {A.}~\bibnamefont
  {Farhan}}, \bibinfo {author} {\bibfnamefont {P.~M.}\ \bibnamefont {Derlet}},
  \bibinfo {author} {\bibfnamefont {A.}~\bibnamefont {Kleibert}}, \bibinfo
  {author} {\bibfnamefont {A.}~\bibnamefont {Balan}}, \bibinfo {author}
  {\bibfnamefont {R.~V.}\ \bibnamefont {Chopdekar}}, \bibinfo {author}
  {\bibfnamefont {M.}~\bibnamefont {Wyss}}, \bibinfo {author} {\bibfnamefont
  {J.}~\bibnamefont {Perron}}, \bibinfo {author} {\bibfnamefont
  {A.}~\bibnamefont {Scholl}}, \bibinfo {author} {\bibfnamefont
  {F.}~\bibnamefont {Nolting}}, \ and\ \bibinfo {author} {\bibfnamefont
  {L.~J.}\ \bibnamefont {Heyderman}},\ }\href@noop {} {\bibfield  {journal}
  {\bibinfo  {journal} {Physical Review Letters}\ }\textbf {\bibinfo {volume}
  {111}},\ \bibinfo {pages} {057204} (\bibinfo {year}
  {2013}{\natexlab{b}})}\BibitemShut {NoStop}%
\bibitem [{\citenamefont {Anghinolfi}\ \emph {et~al.}(2015)\citenamefont
  {Anghinolfi}, \citenamefont {Luetkens}, \citenamefont {Perron}, \citenamefont
  {Flokstra}, \citenamefont {Sendetskyi}, \citenamefont {Suter}, \citenamefont
  {Prokscha}, \citenamefont {Derlet}, \citenamefont {Lee},\ and\ \citenamefont
  {Heyderman}}]{anghinolfi2015}%
  \BibitemOpen
  \bibfield  {author} {\bibinfo {author} {\bibfnamefont {L.}~\bibnamefont
  {Anghinolfi}}, \bibinfo {author} {\bibfnamefont {H.}~\bibnamefont
  {Luetkens}}, \bibinfo {author} {\bibfnamefont {J.}~\bibnamefont {Perron}},
  \bibinfo {author} {\bibfnamefont {M.~G.}\ \bibnamefont {Flokstra}}, \bibinfo
  {author} {\bibfnamefont {O.}~\bibnamefont {Sendetskyi}}, \bibinfo {author}
  {\bibfnamefont {A.}~\bibnamefont {Suter}}, \bibinfo {author} {\bibfnamefont
  {T.}~\bibnamefont {Prokscha}}, \bibinfo {author} {\bibfnamefont {P.~M.}\
  \bibnamefont {Derlet}}, \bibinfo {author} {\bibfnamefont {S.}~\bibnamefont
  {Lee}}, \ and\ \bibinfo {author} {\bibfnamefont {L.~J.}\ \bibnamefont
  {Heyderman}},\ }\href@noop {} {\bibfield  {journal} {\bibinfo  {journal}
  {Nature Communications}\ }\textbf {\bibinfo {volume} {6}} (\bibinfo {year}
  {2015})}\BibitemShut {NoStop}%
\bibitem [{\citenamefont {Kapaklis}\ \emph {et~al.}(2014)\citenamefont
  {Kapaklis}, \citenamefont {Arnalds}, \citenamefont {Farhan}, \citenamefont
  {Chopdekar}, \citenamefont {Balan}, \citenamefont {Scholl}, \citenamefont
  {Heyderman},\ and\ \citenamefont {Hj{\"o}rvarsson}}]{kapaklis2014}%
  \BibitemOpen
  \bibfield  {author} {\bibinfo {author} {\bibfnamefont {V.}~\bibnamefont
  {Kapaklis}}, \bibinfo {author} {\bibfnamefont {U.~B.}\ \bibnamefont
  {Arnalds}}, \bibinfo {author} {\bibfnamefont {A.}~\bibnamefont {Farhan}},
  \bibinfo {author} {\bibfnamefont {R.~V.}\ \bibnamefont {Chopdekar}}, \bibinfo
  {author} {\bibfnamefont {A.}~\bibnamefont {Balan}}, \bibinfo {author}
  {\bibfnamefont {A.}~\bibnamefont {Scholl}}, \bibinfo {author} {\bibfnamefont
  {L.~J.}\ \bibnamefont {Heyderman}}, \ and\ \bibinfo {author} {\bibfnamefont
  {B.}~\bibnamefont {Hj{\"o}rvarsson}},\ }\href@noop {} {\bibfield  {journal}
  {\bibinfo  {journal} {Nature Nanotechnology}\ }\textbf {\bibinfo {volume}
  {9}},\ \bibinfo {pages} {514} (\bibinfo {year} {2014})}\BibitemShut {NoStop}%
\bibitem [{\citenamefont {Morgan}\ \emph {et~al.}(2011)\citenamefont {Morgan},
  \citenamefont {Stein}, \citenamefont {Langridge},\ and\ \citenamefont
  {Marrows}}]{morgan2011}%
  \BibitemOpen
  \bibfield  {author} {\bibinfo {author} {\bibfnamefont {J.~P.}\ \bibnamefont
  {Morgan}}, \bibinfo {author} {\bibfnamefont {A.}~\bibnamefont {Stein}},
  \bibinfo {author} {\bibfnamefont {S.}~\bibnamefont {Langridge}}, \ and\
  \bibinfo {author} {\bibfnamefont {C.~H.}\ \bibnamefont {Marrows}},\
  }\href@noop {} {\bibfield  {journal} {\bibinfo  {journal} {Nature Physics}\
  }\textbf {\bibinfo {volume} {7}},\ \bibinfo {pages} {75} (\bibinfo {year}
  {2011})}\BibitemShut {NoStop}%
\bibitem [{\citenamefont {Chumak}\ \emph {et~al.}(2015)\citenamefont {Chumak},
  \citenamefont {Vasyuchka}, \citenamefont {Serga},\ and\ \citenamefont
  {Hillebrands}}]{chumak2015}%
  \BibitemOpen
  \bibfield  {author} {\bibinfo {author} {\bibfnamefont {A.~V.}\ \bibnamefont
  {Chumak}}, \bibinfo {author} {\bibfnamefont {V.~I.}\ \bibnamefont
  {Vasyuchka}}, \bibinfo {author} {\bibfnamefont {A.~A.}\ \bibnamefont
  {Serga}}, \ and\ \bibinfo {author} {\bibfnamefont {B.}~\bibnamefont
  {Hillebrands}},\ }\href@noop {} {\bibfield  {journal} {\bibinfo  {journal}
  {Nature Physics}\ }\textbf {\bibinfo {volume} {11}},\ \bibinfo {pages} {453}
  (\bibinfo {year} {2015})}\BibitemShut {NoStop}%
\bibitem [{\citenamefont {Farhan}\ \emph {et~al.}(2014)\citenamefont {Farhan},
  \citenamefont {Kleibert}, \citenamefont {Derlet}, \citenamefont {Anghinolfi},
  \citenamefont {Balan}, \citenamefont {Chopdekar}, \citenamefont {Wyss},
  \citenamefont {Gliga}, \citenamefont {Nolting},\ and\ \citenamefont
  {Heyderman}}]{farhan2014}%
  \BibitemOpen
  \bibfield  {author} {\bibinfo {author} {\bibfnamefont {A.}~\bibnamefont
  {Farhan}}, \bibinfo {author} {\bibfnamefont {A.}~\bibnamefont {Kleibert}},
  \bibinfo {author} {\bibfnamefont {P.~M.}\ \bibnamefont {Derlet}}, \bibinfo
  {author} {\bibfnamefont {L.}~\bibnamefont {Anghinolfi}}, \bibinfo {author}
  {\bibfnamefont {A.}~\bibnamefont {Balan}}, \bibinfo {author} {\bibfnamefont
  {R.~V.}\ \bibnamefont {Chopdekar}}, \bibinfo {author} {\bibfnamefont
  {M.}~\bibnamefont {Wyss}}, \bibinfo {author} {\bibfnamefont {S.}~\bibnamefont
  {Gliga}}, \bibinfo {author} {\bibfnamefont {F.}~\bibnamefont {Nolting}}, \
  and\ \bibinfo {author} {\bibfnamefont {L.~J.}\ \bibnamefont {Heyderman}},\
  }\href@noop {} {\bibfield  {journal} {\bibinfo  {journal} {Physical Review
  B}\ }\textbf {\bibinfo {volume} {89}},\ \bibinfo {pages} {214405} (\bibinfo
  {year} {2014})}\BibitemShut {NoStop}%
\bibitem [{\citenamefont {M{\"o}ller}\ and\ \citenamefont
  {Moessner}(2009)}]{moller2009}%
  \BibitemOpen
  \bibfield  {author} {\bibinfo {author} {\bibfnamefont {G.}~\bibnamefont
  {M{\"o}ller}}\ and\ \bibinfo {author} {\bibfnamefont {R.}~\bibnamefont
  {Moessner}},\ }\href@noop {} {\bibfield  {journal} {\bibinfo  {journal}
  {Physical Review B}\ }\textbf {\bibinfo {volume} {80}},\ \bibinfo {pages}
  {140409} (\bibinfo {year} {2009})}\BibitemShut {NoStop}%
\bibitem [{\citenamefont {Chern}\ \emph {et~al.}(2011)\citenamefont {Chern},
  \citenamefont {Mellado},\ and\ \citenamefont {Tchernyshyov}}]{chern2011}%
  \BibitemOpen
  \bibfield  {author} {\bibinfo {author} {\bibfnamefont {G.-W.}\ \bibnamefont
  {Chern}}, \bibinfo {author} {\bibfnamefont {P.}~\bibnamefont {Mellado}}, \
  and\ \bibinfo {author} {\bibfnamefont {O.}~\bibnamefont {Tchernyshyov}},\
  }\href@noop {} {\bibfield  {journal} {\bibinfo  {journal} {Physical Review
  Letters}\ }\textbf {\bibinfo {volume} {106}},\ \bibinfo {pages} {207202}
  (\bibinfo {year} {2011})}\BibitemShut {NoStop}%
\bibitem [{\citenamefont {Krawczyk}\ and\ \citenamefont
  {Grundler}(2014)}]{krawczyk2014}%
  \BibitemOpen
  \bibfield  {author} {\bibinfo {author} {\bibfnamefont {M.}~\bibnamefont
  {Krawczyk}}\ and\ \bibinfo {author} {\bibfnamefont {D.}~\bibnamefont
  {Grundler}},\ }\href@noop {} {\bibfield  {journal} {\bibinfo  {journal}
  {Journal of Physics: Condensed Matter}\ }\textbf {\bibinfo {volume} {26}},\
  \bibinfo {pages} {123202} (\bibinfo {year} {2014})}\BibitemShut {NoStop}%
\bibitem [{\citenamefont {Morgan}\ \emph {et~al.}(2012)\citenamefont {Morgan},
  \citenamefont {Kinane}, \citenamefont {Charlton}, \citenamefont {Stein},
  \citenamefont {S{\'a}nchez-Hanke}, \citenamefont {Arena}, \citenamefont
  {Langridge},\ and\ \citenamefont {Marrows}}]{morgan2012}%
  \BibitemOpen
  \bibfield  {author} {\bibinfo {author} {\bibfnamefont {J.~P.}\ \bibnamefont
  {Morgan}}, \bibinfo {author} {\bibfnamefont {C.~J.}\ \bibnamefont {Kinane}},
  \bibinfo {author} {\bibfnamefont {T.~R.}\ \bibnamefont {Charlton}}, \bibinfo
  {author} {\bibfnamefont {A.}~\bibnamefont {Stein}}, \bibinfo {author}
  {\bibfnamefont {C.}~\bibnamefont {S{\'a}nchez-Hanke}}, \bibinfo {author}
  {\bibfnamefont {D.~A.}\ \bibnamefont {Arena}}, \bibinfo {author}
  {\bibfnamefont {S.}~\bibnamefont {Langridge}}, \ and\ \bibinfo {author}
  {\bibfnamefont {C.~H.}\ \bibnamefont {Marrows}},\ }\href@noop {} {\bibfield
  {journal} {\bibinfo  {journal} {AIP Adv}\ }\textbf {\bibinfo {volume} {2}},\
  \bibinfo {pages} {022163} (\bibinfo {year} {2012})}\BibitemShut {NoStop}%
\bibitem [{\citenamefont {Perron}\ \emph {et~al.}(2013)\citenamefont {Perron},
  \citenamefont {Anghinolfi}, \citenamefont {Tudu}, \citenamefont {Jaouen},
  \citenamefont {Tonnerre}, \citenamefont {Sacchi}, \citenamefont {Nolting},
  \citenamefont {L{\"u}ning},\ and\ \citenamefont {Heyderman}}]{perron2013}%
  \BibitemOpen
  \bibfield  {author} {\bibinfo {author} {\bibfnamefont {J.}~\bibnamefont
  {Perron}}, \bibinfo {author} {\bibfnamefont {L.}~\bibnamefont {Anghinolfi}},
  \bibinfo {author} {\bibfnamefont {B.}~\bibnamefont {Tudu}}, \bibinfo {author}
  {\bibfnamefont {N.}~\bibnamefont {Jaouen}}, \bibinfo {author} {\bibfnamefont
  {J.-M.}\ \bibnamefont {Tonnerre}}, \bibinfo {author} {\bibfnamefont
  {M.}~\bibnamefont {Sacchi}}, \bibinfo {author} {\bibfnamefont
  {F.}~\bibnamefont {Nolting}}, \bibinfo {author} {\bibfnamefont
  {J.}~\bibnamefont {L{\"u}ning}}, \ and\ \bibinfo {author} {\bibfnamefont
  {L.~J.}\ \bibnamefont {Heyderman}},\ }\href@noop {} {\bibfield  {journal}
  {\bibinfo  {journal} {Physical Review B}\ }\textbf {\bibinfo {volume} {88}},\
  \bibinfo {pages} {214424} (\bibinfo {year} {2013})}\BibitemShut {NoStop}%
\bibitem [{\citenamefont {Arnalds}\ \emph {et~al.}(2012)\citenamefont
  {Arnalds}, \citenamefont {Hase}, \citenamefont {Papaioannou}, \citenamefont
  {Raanaei}, \citenamefont {Abrudan}, \citenamefont {Charlton}, \citenamefont
  {Langridge},\ and\ \citenamefont {Hj{\"o}rvarsson}}]{arnalds2012}%
  \BibitemOpen
  \bibfield  {author} {\bibinfo {author} {\bibfnamefont {U.~B.}\ \bibnamefont
  {Arnalds}}, \bibinfo {author} {\bibfnamefont {T.~P.~A.}\ \bibnamefont
  {Hase}}, \bibinfo {author} {\bibfnamefont {E.~T.}\ \bibnamefont
  {Papaioannou}}, \bibinfo {author} {\bibfnamefont {H.}~\bibnamefont
  {Raanaei}}, \bibinfo {author} {\bibfnamefont {R.}~\bibnamefont {Abrudan}},
  \bibinfo {author} {\bibfnamefont {T.~R.}\ \bibnamefont {Charlton}}, \bibinfo
  {author} {\bibfnamefont {S.}~\bibnamefont {Langridge}}, \ and\ \bibinfo
  {author} {\bibfnamefont {B.}~\bibnamefont {Hj{\"o}rvarsson}},\ }\href@noop {}
  {\bibfield  {journal} {\bibinfo  {journal} {Physical Review B}\ }\textbf
  {\bibinfo {volume} {86}},\ \bibinfo {pages} {064426} (\bibinfo {year}
  {2012})}\BibitemShut {NoStop}%
\bibitem [{\citenamefont {Tabata}\ \emph {et~al.}(2006)\citenamefont {Tabata},
  \citenamefont {Kadowaki}, \citenamefont {Matsuhira}, \citenamefont {Hiroi},
  \citenamefont {Aso}, \citenamefont {Ressouche},\ and\ \citenamefont
  {F{\aa}k}}]{tabata2006}%
  \BibitemOpen
  \bibfield  {author} {\bibinfo {author} {\bibfnamefont {Y.}~\bibnamefont
  {Tabata}}, \bibinfo {author} {\bibfnamefont {H.}~\bibnamefont {Kadowaki}},
  \bibinfo {author} {\bibfnamefont {K.}~\bibnamefont {Matsuhira}}, \bibinfo
  {author} {\bibfnamefont {Z.}~\bibnamefont {Hiroi}}, \bibinfo {author}
  {\bibfnamefont {N.}~\bibnamefont {Aso}}, \bibinfo {author} {\bibfnamefont
  {E.}~\bibnamefont {Ressouche}}, \ and\ \bibinfo {author} {\bibfnamefont
  {B.}~\bibnamefont {F{\aa}k}},\ }\href@noop {} {\bibfield  {journal} {\bibinfo
   {journal} {Physical Review Letters}\ }\textbf {\bibinfo {volume} {97}},\
  \bibinfo {pages} {257205} (\bibinfo {year} {2006})}\BibitemShut {NoStop}%
\bibitem [{\citenamefont {Fennell}\ \emph {et~al.}(2007)\citenamefont
  {Fennell}, \citenamefont {Bramwell}, \citenamefont {McMorrow}, \citenamefont
  {Manuel},\ and\ \citenamefont {Wildes}}]{fennell2007}%
  \BibitemOpen
  \bibfield  {author} {\bibinfo {author} {\bibfnamefont {T.}~\bibnamefont
  {Fennell}}, \bibinfo {author} {\bibfnamefont {S.~T.}\ \bibnamefont
  {Bramwell}}, \bibinfo {author} {\bibfnamefont {D.~F.}\ \bibnamefont
  {McMorrow}}, \bibinfo {author} {\bibfnamefont {P.}~\bibnamefont {Manuel}}, \
  and\ \bibinfo {author} {\bibfnamefont {A.~R.}\ \bibnamefont {Wildes}},\
  }\href@noop {} {\bibfield  {journal} {\bibinfo  {journal} {Nature Physics}\
  }\textbf {\bibinfo {volume} {3}},\ \bibinfo {pages} {566} (\bibinfo {year}
  {2007})}\BibitemShut {NoStop}%
\bibitem [{\citenamefont {Fennell}\ \emph {et~al.}(2009)\citenamefont
  {Fennell}, \citenamefont {Deen}, \citenamefont {Wildes}, \citenamefont
  {Schmalzl}, \citenamefont {Prabhakaran}, \citenamefont {Boothroyd},
  \citenamefont {Aldus}, \citenamefont {McMorrow},\ and\ \citenamefont
  {Bramwell}}]{fennell2009}%
  \BibitemOpen
  \bibfield  {author} {\bibinfo {author} {\bibfnamefont {T.}~\bibnamefont
  {Fennell}}, \bibinfo {author} {\bibfnamefont {P.~P.}\ \bibnamefont {Deen}},
  \bibinfo {author} {\bibfnamefont {A.~R.}\ \bibnamefont {Wildes}}, \bibinfo
  {author} {\bibfnamefont {K.}~\bibnamefont {Schmalzl}}, \bibinfo {author}
  {\bibfnamefont {D.}~\bibnamefont {Prabhakaran}}, \bibinfo {author}
  {\bibfnamefont {A.~T.}\ \bibnamefont {Boothroyd}}, \bibinfo {author}
  {\bibfnamefont {R.~J.}\ \bibnamefont {Aldus}}, \bibinfo {author}
  {\bibfnamefont {D.~F.}\ \bibnamefont {McMorrow}}, \ and\ \bibinfo {author}
  {\bibfnamefont {S.~T.}\ \bibnamefont {Bramwell}},\ }\href@noop {} {\bibfield
  {journal} {\bibinfo  {journal} {Science}\ }\textbf {\bibinfo {volume}
  {326}},\ \bibinfo {pages} {415} (\bibinfo {year} {2009})}\BibitemShut
  {NoStop}%
\bibitem [{\citenamefont {Brooks-Bartlett}\ \emph {et~al.}(2014)\citenamefont
  {Brooks-Bartlett}, \citenamefont {Banks}, \citenamefont {Jaubert},
  \citenamefont {Harman-Clarke},\ and\ \citenamefont
  {Holdsworth}}]{brooksbartlett2014}%
  \BibitemOpen
  \bibfield  {author} {\bibinfo {author} {\bibfnamefont {M.~E.}\ \bibnamefont
  {Brooks-Bartlett}}, \bibinfo {author} {\bibfnamefont {S.~T.}\ \bibnamefont
  {Banks}}, \bibinfo {author} {\bibfnamefont {L.~D.~C.}\ \bibnamefont
  {Jaubert}}, \bibinfo {author} {\bibfnamefont {A.}~\bibnamefont
  {Harman-Clarke}}, \ and\ \bibinfo {author} {\bibfnamefont {P.~C.~W.}\
  \bibnamefont {Holdsworth}},\ }\href@noop {} {\bibfield  {journal} {\bibinfo
  {journal} {Phys. Rev. X}\ }\textbf {\bibinfo {volume} {4}},\ \bibinfo {pages}
  {011007} (\bibinfo {year} {2014})}\BibitemShut {NoStop}%
\bibitem [{\citenamefont {M{\"o}ller}\ and\ \citenamefont
  {Moessner}(2006)}]{moller2006}%
  \BibitemOpen
  \bibfield  {author} {\bibinfo {author} {\bibfnamefont {G.}~\bibnamefont
  {M{\"o}ller}}\ and\ \bibinfo {author} {\bibfnamefont {R.}~\bibnamefont
  {Moessner}},\ }\href@noop {} {\bibfield  {journal} {\bibinfo  {journal}
  {Physical Review Letters}\ }\textbf {\bibinfo {volume} {96}},\ \bibinfo
  {pages} {237202} (\bibinfo {year} {2006})}\BibitemShut {NoStop}%
\bibitem [{\citenamefont {Staub}\ \emph {et~al.}(2008)\citenamefont {Staub},
  \citenamefont {Scagnoli}, \citenamefont {Bodenthin}, \citenamefont
  {Garcia-Fernandez}, \citenamefont {Wetter}, \citenamefont {Mulders},
  \citenamefont {Grimmer},\ and\ \citenamefont {Horisberger}}]{staub2008}%
  \BibitemOpen
  \bibfield  {author} {\bibinfo {author} {\bibfnamefont {U.}~\bibnamefont
  {Staub}}, \bibinfo {author} {\bibfnamefont {V.}~\bibnamefont {Scagnoli}},
  \bibinfo {author} {\bibfnamefont {Y.}~\bibnamefont {Bodenthin}}, \bibinfo
  {author} {\bibfnamefont {M.}~\bibnamefont {Garcia-Fernandez}}, \bibinfo
  {author} {\bibfnamefont {R.}~\bibnamefont {Wetter}}, \bibinfo {author}
  {\bibfnamefont {A.~M.}\ \bibnamefont {Mulders}}, \bibinfo {author}
  {\bibfnamefont {H.}~\bibnamefont {Grimmer}}, \ and\ \bibinfo {author}
  {\bibfnamefont {M.}~\bibnamefont {Horisberger}},\ }\href@noop {} {\bibfield
  {journal} {\bibinfo  {journal} {Journal of Synchrotron Radiation}\ }\textbf
  {\bibinfo {volume} {15}},\ \bibinfo {pages} {469} (\bibinfo {year}
  {2008})}\BibitemShut {NoStop}%
\bibitem [{\citenamefont {Flechsig}\ \emph {et~al.}(2010)\citenamefont
  {Flechsig}, \citenamefont {Nolting}, \citenamefont {Fraile~Rodr\'{\i}guez},
  \citenamefont {Krempasky}, \citenamefont {Quitmann}, \citenamefont {Schmidt},
  \citenamefont {Spielmann}, \citenamefont {Zimoch}, \citenamefont {Garrett},
  \citenamefont {Gentle} \emph {et~al.}}]{flechsig2010}%
  \BibitemOpen
  \bibfield  {author} {\bibinfo {author} {\bibfnamefont {U.}~\bibnamefont
  {Flechsig}}, \bibinfo {author} {\bibfnamefont {F.}~\bibnamefont {Nolting}},
  \bibinfo {author} {\bibfnamefont {A.}~\bibnamefont {Fraile~Rodr\'{\i}guez}},
  \bibinfo {author} {\bibfnamefont {J.}~\bibnamefont {Krempasky}}, \bibinfo
  {author} {\bibfnamefont {C.}~\bibnamefont {Quitmann}}, \bibinfo {author}
  {\bibfnamefont {T.}~\bibnamefont {Schmidt}}, \bibinfo {author} {\bibfnamefont
  {S.}~\bibnamefont {Spielmann}}, \bibinfo {author} {\bibfnamefont
  {D.}~\bibnamefont {Zimoch}}, \bibinfo {author} {\bibfnamefont
  {R.}~\bibnamefont {Garrett}}, \bibinfo {author} {\bibfnamefont
  {I.}~\bibnamefont {Gentle}},  \emph {et~al.},\ }in\ \href@noop {} {\emph
  {\bibinfo {booktitle} {AIP Conference Proceedings}}},\ Vol.\ \bibinfo
  {volume} {1234}\ (\bibinfo {year} {2010})\ p.\ \bibinfo {pages}
  {319}\BibitemShut {NoStop}%
\bibitem [{\citenamefont {Kraft}\ \emph {et~al.}(2009)\citenamefont {Kraft},
  \citenamefont {Bergamaschi}, \citenamefont {Broennimann}, \citenamefont
  {Dinapoli}, \citenamefont {Eikenberry}, \citenamefont {Henrich},
  \citenamefont {Johnson}, \citenamefont {Mozzanica}, \citenamefont
  {Schlep{\"u}tz}, \citenamefont {Willmott} \emph {et~al.}}]{kraft2009}%
  \BibitemOpen
  \bibfield  {author} {\bibinfo {author} {\bibfnamefont {P.}~\bibnamefont
  {Kraft}}, \bibinfo {author} {\bibfnamefont {A.}~\bibnamefont {Bergamaschi}},
  \bibinfo {author} {\bibfnamefont {C.}~\bibnamefont {Broennimann}}, \bibinfo
  {author} {\bibfnamefont {R.}~\bibnamefont {Dinapoli}}, \bibinfo {author}
  {\bibfnamefont {E.~F.}\ \bibnamefont {Eikenberry}}, \bibinfo {author}
  {\bibfnamefont {B.}~\bibnamefont {Henrich}}, \bibinfo {author} {\bibfnamefont
  {I.}~\bibnamefont {Johnson}}, \bibinfo {author} {\bibfnamefont
  {A.}~\bibnamefont {Mozzanica}}, \bibinfo {author} {\bibfnamefont {C.~M.}\
  \bibnamefont {Schlep{\"u}tz}}, \bibinfo {author} {\bibfnamefont {P.~R.}\
  \bibnamefont {Willmott}},  \emph {et~al.},\ }\href@noop {} {\bibfield
  {journal} {\bibinfo  {journal} {Journal of Synchrotron Radiation}\ }\textbf
  {\bibinfo {volume} {16}},\ \bibinfo {pages} {368} (\bibinfo {year}
  {2009})}\BibitemShut {NoStop}%
\bibitem [{\citenamefont {Blume}(1985)}]{blume1985}%
  \BibitemOpen
  \bibfield  {author} {\bibinfo {author} {\bibfnamefont {M.}~\bibnamefont
  {Blume}},\ }\href@noop {} {\bibfield  {journal} {\bibinfo  {journal} {Journal
  of Applied Physics}\ }\textbf {\bibinfo {volume} {57}},\ \bibinfo {pages}
  {3615} (\bibinfo {year} {1985})}\BibitemShut {NoStop}%
\bibitem [{\citenamefont {Hannon}\ \emph {et~al.}(1989)\citenamefont {Hannon},
  \citenamefont {Trammell}, \citenamefont {Blume},\ and\ \citenamefont
  {Gibbs}}]{hannon1989}%
  \BibitemOpen
  \bibfield  {author} {\bibinfo {author} {\bibfnamefont {J.~P.}\ \bibnamefont
  {Hannon}}, \bibinfo {author} {\bibfnamefont {G.~T.}\ \bibnamefont
  {Trammell}}, \bibinfo {author} {\bibfnamefont {M.}~\bibnamefont {Blume}}, \
  and\ \bibinfo {author} {\bibfnamefont {D.}~\bibnamefont {Gibbs}},\
  }\href@noop {} {\bibfield  {journal} {\bibinfo  {journal} {Physical Review
  Letters}\ }\textbf {\bibinfo {volume} {62}},\ \bibinfo {pages} {2644}
  (\bibinfo {year} {1989})}\BibitemShut {NoStop}%
\bibitem [{\citenamefont {van~der Laan}(2008)}]{van2008}%
  \BibitemOpen
  \bibfield  {author} {\bibinfo {author} {\bibfnamefont {G.}~\bibnamefont
  {van~der Laan}},\ }\href@noop {} {\bibfield  {journal} {\bibinfo  {journal}
  {Comptes Rendus Physique}\ }\textbf {\bibinfo {volume} {9}},\ \bibinfo
  {pages} {570} (\bibinfo {year} {2008})}\BibitemShut {NoStop}%
\bibitem [{\citenamefont {Als-Nielsen}\ and\ \citenamefont
  {McMorrow}(2011)}]{als2011}%
  \BibitemOpen
  \bibfield  {author} {\bibinfo {author} {\bibfnamefont {J.}~\bibnamefont
  {Als-Nielsen}}\ and\ \bibinfo {author} {\bibfnamefont {D.}~\bibnamefont
  {McMorrow}},\ }\href@noop {} {\emph {\bibinfo {title} {Elements of modern
  X-ray physics}}}\ (\bibinfo  {publisher} {John Wiley \& Sons},\ \bibinfo
  {year} {2011})\BibitemShut {NoStop}%
\bibitem [{\citenamefont {Balents}(2010)}]{bal2010}%
  \BibitemOpen
  \bibfield  {author} {\bibinfo {author} {\bibfnamefont {L.}~\bibnamefont
  {Balents}},\ }\href@noop {} {\bibfield  {journal} {\bibinfo  {journal}
  {Nature}\ }\textbf {\bibinfo {volume} {464}},\ \bibinfo {pages} {199}
  (\bibinfo {year} {2010})}\BibitemShut {NoStop}%
\bibitem [{\citenamefont {Villain}(1979)}]{villain1979}%
  \BibitemOpen
  \bibfield  {author} {\bibinfo {author} {\bibfnamefont {J.}~\bibnamefont
  {Villain}},\ }\href@noop {} {\bibfield  {journal} {\bibinfo  {journal}
  {Zeitschrift f{\"u}r Physik B Condensed Matter}\ }\textbf {\bibinfo {volume}
  {33}},\ \bibinfo {pages} {31} (\bibinfo {year} {1979})}\BibitemShut {NoStop}%
\bibitem [{\citenamefont {Isakov}\ \emph {et~al.}(2004)\citenamefont {Isakov},
  \citenamefont {Gregor}, \citenamefont {Moessner},\ and\ \citenamefont
  {Sondhi}}]{isakov2004}%
  \BibitemOpen
  \bibfield  {author} {\bibinfo {author} {\bibfnamefont {S.~V.}\ \bibnamefont
  {Isakov}}, \bibinfo {author} {\bibfnamefont {K.}~\bibnamefont {Gregor}},
  \bibinfo {author} {\bibfnamefont {R.}~\bibnamefont {Moessner}}, \ and\
  \bibinfo {author} {\bibfnamefont {S.~L.}\ \bibnamefont {Sondhi}},\
  }\href@noop {} {\bibfield  {journal} {\bibinfo  {journal} {Physical Review
  Letters}\ }\textbf {\bibinfo {volume} {93}},\ \bibinfo {pages} {167204}
  (\bibinfo {year} {2004})}\BibitemShut {NoStop}%
\bibitem [{\citenamefont {Moessner}\ and\ \citenamefont
  {Sondhi}(2003)}]{moessner2003}%
  \BibitemOpen
  \bibfield  {author} {\bibinfo {author} {\bibfnamefont {R.}~\bibnamefont
  {Moessner}}\ and\ \bibinfo {author} {\bibfnamefont {S.~L.}~\bibnamefont
  {Sondhi}},\ }\href@noop {} {\bibfield  {journal} {\bibinfo  {journal}
  {Physical Review B}\ }\textbf {\bibinfo {volume} {68}},\ \bibinfo {pages}
  {064411} (\bibinfo {year} {2003})}\BibitemShut {NoStop}%
\bibitem [{\citenamefont {Skalyo~Jr}\ \emph {et~al.}(1970)\citenamefont
  {Skalyo~Jr}, \citenamefont {Frazer},\ and\ \citenamefont
  {Shirane}}]{skalyo1970}%
  \BibitemOpen
  \bibfield  {author} {\bibinfo {author} {\bibfnamefont {J.}~\bibnamefont
  {Skalyo~Jr}}, \bibinfo {author} {\bibfnamefont {B.~C.}\ \bibnamefont
  {Frazer}}, \ and\ \bibinfo {author} {\bibfnamefont {G.}~\bibnamefont
  {Shirane}},\ }\href@noop {} {\bibfield  {journal} {\bibinfo  {journal}
  {Physical Review B}\ }\textbf {\bibinfo {volume} {1}},\ \bibinfo {pages}
  {278} (\bibinfo {year} {1970})}\BibitemShut {NoStop}%
\bibitem [{\citenamefont {Youngblood}\ and\ \citenamefont
  {Axe}(1981)}]{youngblood1981}%
  \BibitemOpen
  \bibfield  {author} {\bibinfo {author} {\bibfnamefont {R.~W.}\ \bibnamefont
  {Youngblood}}\ and\ \bibinfo {author} {\bibfnamefont {J.~D.}\ \bibnamefont
  {Axe}},\ }\href@noop {} {\bibfield  {journal} {\bibinfo  {journal} {Physical
  Review B}\ }\textbf {\bibinfo {volume} {23}},\ \bibinfo {pages} {232}
  (\bibinfo {year} {1981})}\BibitemShut {NoStop}%
\bibitem [{\citenamefont {Henley}(2005)}]{henley2005}%
  \BibitemOpen
  \bibfield  {author} {\bibinfo {author} {\bibfnamefont {C.~L.}\ \bibnamefont
  {Henley}},\ }\href@noop {} {\bibfield  {journal} {\bibinfo  {journal}
  {Physical Review B}\ }\textbf {\bibinfo {volume} {71}},\ \bibinfo {pages}
  {014424} (\bibinfo {year} {2005})}\BibitemShut {NoStop}%
\bibitem [{\citenamefont {Anderson}(1956)}]{anderson1956}%
  \BibitemOpen
  \bibfield  {author} {\bibinfo {author} {\bibfnamefont {P.~W.}\ \bibnamefont
  {Anderson}},\ }\href@noop {} {\bibfield  {journal} {\bibinfo  {journal}
  {Physical Review}\ }\textbf {\bibinfo {volume} {102}},\ \bibinfo {pages}
  {1008} (\bibinfo {year} {1956})}\BibitemShut {NoStop}%
\bibitem [{\citenamefont {Burzlaff}\ and\ \citenamefont
  {Zimmermann}(2006)}]{burzlaff2006}%
  \BibitemOpen
  \bibfield  {author} {\bibinfo {author} {\bibfnamefont {H.}~\bibnamefont
  {Burzlaff}}\ and\ \bibinfo {author} {\bibfnamefont {H.}~\bibnamefont
  {Zimmermann}},\ }in\ \href@noop {} {\emph {\bibinfo {booktitle}
  {International Tables for Crystallography Volume A: Space-group symmetry}}}\
  (\bibinfo  {publisher} {Springer},\ \bibinfo {year} {2006})\ pp.\ \bibinfo
  {pages} {742--749}\BibitemShut {NoStop}%
\bibitem [{\citenamefont {Steurer}\ and\ \citenamefont
  {Deloudi}(2009)}]{steurer2009}%
  \BibitemOpen
  \bibfield  {author} {\bibinfo {author} {\bibfnamefont {W.}~\bibnamefont
  {Steurer}}\ and\ \bibinfo {author} {\bibfnamefont {S.}~\bibnamefont
  {Deloudi}},\ }\href@noop {} {\emph {\bibinfo {title} {Crystallography of
  quasicrystals: concepts, methods and structures}}},\ Vol.\ \bibinfo {volume}
  {126}\ (\bibinfo  {publisher} {Springer Science \& Business Media},\ \bibinfo
  {year} {2009})\ pp.\ \bibinfo {pages} {18--20}\BibitemShut {NoStop}%
\bibitem [{\citenamefont {Huse}\ \emph {et~al.}(2003)\citenamefont {Huse},
  \citenamefont {Krauth}, \citenamefont {Moessner},\ and\ \citenamefont
  {Sondhi}}]{huse2003}%
  \BibitemOpen
  \bibfield  {author} {\bibinfo {author} {\bibfnamefont {D.~A.}\ \bibnamefont
  {Huse}}, \bibinfo {author} {\bibfnamefont {W.}~\bibnamefont {Krauth}},
  \bibinfo {author} {\bibfnamefont {R.}~\bibnamefont {Moessner}}, \ and\
  \bibinfo {author} {\bibfnamefont {S.~L.}\ \bibnamefont {Sondhi}},\
  }\href@noop {} {\bibfield  {journal} {\bibinfo  {journal} {Physical Review
  Letters}\ }\textbf {\bibinfo {volume} {91}},\ \bibinfo {pages} {167004}
  (\bibinfo {year} {2003})}\BibitemShut {NoStop}%
\bibitem [{\citenamefont {Henley}(2010)}]{henley2010}%
  \BibitemOpen
  \bibfield  {author} {\bibinfo {author} {\bibfnamefont {C.~L.}\ \bibnamefont
  {Henley}},\ }\href@noop {} {\bibfield  {journal} {\bibinfo  {journal} {Annu.
  Rev. Condens. Matter Phys.}\ }\textbf {\bibinfo {volume} {1}},\ \bibinfo
  {pages} {179} (\bibinfo {year} {2010})}\BibitemShut {NoStop}%
\bibitem [{\citenamefont {Udagawa}\ \emph {et~al.}(2002)\citenamefont
  {Udagawa}, \citenamefont {Ogata},\ and\ \citenamefont {Hiroi}}]{udagawa2002}%
  \BibitemOpen
  \bibfield  {author} {\bibinfo {author} {\bibfnamefont {M.}~\bibnamefont
  {Udagawa}}, \bibinfo {author} {\bibfnamefont {M.}~\bibnamefont {Ogata}}, \
  and\ \bibinfo {author} {\bibfnamefont {Z.}~\bibnamefont {Hiroi}},\
  }\href@noop {} {\bibfield  {journal} {\bibinfo  {journal} {Journal of the
  Physical Society of Japan}\ }\textbf {\bibinfo {volume} {71}},\ \bibinfo
  {pages} {2365} (\bibinfo {year} {2002})}\BibitemShut {NoStop}%
\bibitem [{\citenamefont {Misguich}\ \emph {et~al.}(2002)\citenamefont
  {Misguich}, \citenamefont {Serban},\ and\ \citenamefont
  {Pasquier}}]{misguich2002}%
  \BibitemOpen
  \bibfield  {author} {\bibinfo {author} {\bibfnamefont {G.}~\bibnamefont
  {Misguich}}, \bibinfo {author} {\bibfnamefont {D.}~\bibnamefont {Serban}}, \
  and\ \bibinfo {author} {\bibfnamefont {V.}~\bibnamefont {Pasquier}},\
  }\href@noop {} {\bibfield  {journal} {\bibinfo  {journal} {Physical Review
  Letters}\ }\textbf {\bibinfo {volume} {89}},\ \bibinfo {pages} {137202}
  (\bibinfo {year} {2002})}\BibitemShut {NoStop}%
\bibitem [{Note1()}]{Note1}%
  \BibitemOpen
  \bibinfo {note} {More general flux dimers could be defined, such that the
  dimer carries an effective flux of zero, and the two adjacent vertices have
  one spin in- and out- each. However, one cannot assign such dimers in a
  unique fashion in either of the ice I or ice II phases.}\BibitemShut {Stop}%
\bibitem [{\citenamefont {Carbone}\ \emph {et~al.}(2011)\citenamefont
  {Carbone}, \citenamefont {Gardonio}, \citenamefont {Moras}, \citenamefont
  {Lounis}, \citenamefont {Heide}, \citenamefont {Bihlmayer}, \citenamefont
  {Atodiresei}, \citenamefont {Dederichs}, \citenamefont {Bl{\"u}gel},
  \citenamefont {Vlaic} \emph {et~al.}}]{carbone2011}%
  \BibitemOpen
  \bibfield  {author} {\bibinfo {author} {\bibfnamefont {C.}~\bibnamefont
  {Carbone}}, \bibinfo {author} {\bibfnamefont {S.}~\bibnamefont {Gardonio}},
  \bibinfo {author} {\bibfnamefont {P.}~\bibnamefont {Moras}}, \bibinfo
  {author} {\bibfnamefont {S.}~\bibnamefont {Lounis}}, \bibinfo {author}
  {\bibfnamefont {M.}~\bibnamefont {Heide}}, \bibinfo {author} {\bibfnamefont
  {G.}~\bibnamefont {Bihlmayer}}, \bibinfo {author} {\bibfnamefont
  {N.}~\bibnamefont {Atodiresei}}, \bibinfo {author} {\bibfnamefont {P.~H.}\
  \bibnamefont {Dederichs}}, \bibinfo {author} {\bibfnamefont {S.}~\bibnamefont
  {Bl{\"u}gel}}, \bibinfo {author} {\bibfnamefont {S.}~\bibnamefont {Vlaic}},
  \emph {et~al.},\ }\href@noop {} {\bibfield  {journal} {\bibinfo  {journal}
  {Advanced Functional Materials}\ }\textbf {\bibinfo {volume} {21}},\ \bibinfo
  {pages} {1212} (\bibinfo {year} {2011})}\BibitemShut {NoStop}%
\bibitem [{\citenamefont {M{\"u}hlbauer}\ \emph {et~al.}(2009)\citenamefont
  {M{\"u}hlbauer}, \citenamefont {Binz}, \citenamefont {Jonietz}, \citenamefont
  {Pfleiderer}, \citenamefont {Rosch}, \citenamefont {Neubauer}, \citenamefont
  {Georgii},\ and\ \citenamefont {B{\"o}ni}}]{muhlbauer2009}%
  \BibitemOpen
  \bibfield  {author} {\bibinfo {author} {\bibfnamefont {S.}~\bibnamefont
  {M{\"u}hlbauer}}, \bibinfo {author} {\bibfnamefont {B.}~\bibnamefont {Binz}},
  \bibinfo {author} {\bibfnamefont {F.}~\bibnamefont {Jonietz}}, \bibinfo
  {author} {\bibfnamefont {C.}~\bibnamefont {Pfleiderer}}, \bibinfo {author}
  {\bibfnamefont {A.}~\bibnamefont {Rosch}}, \bibinfo {author} {\bibfnamefont
  {A.}~\bibnamefont {Neubauer}}, \bibinfo {author} {\bibfnamefont
  {R.}~\bibnamefont {Georgii}}, \ and\ \bibinfo {author} {\bibfnamefont
  {P.}~\bibnamefont {B{\"o}ni}},\ }\href@noop {} {\bibfield  {journal}
  {\bibinfo  {journal} {Science}\ }\textbf {\bibinfo {volume} {323}},\ \bibinfo
  {pages} {915} (\bibinfo {year} {2009})}\BibitemShut {NoStop}%
\end{thebibliography}
%
\end{document}